\documentclass[prd,amsmath,amssymb,preprint,numerical]{revtex4-1}

\usepackage[pdftex]{graphicx}
\usepackage[usenames,dvipsnames]{color}
\usepackage{dcolumn}
\usepackage{bm}
\usepackage{subfigure}

\usepackage [pagewise, mathlines, displaymath]{lineno}

\usepackage{boxedminipage}

\begin{document}

\title{Probing the radio emission from air showers with polarization measurements}


\author{A.~Aab}
\affiliation{Universit\"{a}t Siegen, Siegen,
Germany}
\author{P.~Abreu}
\affiliation{LIP and Instituto Superior T\'{e}cnico, Technical
University of Lisbon,
Portugal}
\author{M.~Aglietta}
\affiliation{Osservatorio Astrofisico di Torino  (INAF),
Universit\`{a} di Torino and Sezione INFN, Torino,
Italy}
\author{M.~Ahlers}
\affiliation{University of Wisconsin, Madison, WI,
USA}
\author{E.J.~Ahn}
\affiliation{Fermilab, Batavia, IL,
USA}
\author{I.F.M.~Albuquerque}
\affiliation{Universidade de S\~{a}o Paulo, Instituto de F\'{\i}sica,
S\~{a}o Paulo, SP,
Brazil}
\author{I.~Allekotte}
\affiliation{Centro At\'{o}mico Bariloche and Instituto Balseiro
(CNEA-UNCuyo-CONICET), San Carlos de Bariloche,
Argentina}
\author{J.~Allen}
\affiliation{New York University, New York, NY,
USA}
\author{P.~Allison}
\affiliation{Ohio State University, Columbus, OH,
USA}
\author{A.~Almela}
\affiliation{Universidad Tecnol\'{o}gica Nacional - Facultad
Regional Buenos Aires, Buenos Aires,
Argentina}
\affiliation{Instituto de Tecnolog\'{\i}as en Detecci\'{o}n y
Astropart\'{\i}culas (CNEA, CONICET, UNSAM), Buenos Aires,
Argentina}
% ...
\collaboration{The Pierre Auger Collaboration}
\email{{\tt auger\_spokespersons@fnal.gov}}
\noaffiliation

\date{\today}

\begin{abstract}
The emission of radio waves from air showers has been attributed to the so-called geomagnetic emission process. At frequencies around 50 MHz this process leads to coherent radiation which can be observed with rather simple setups. The direction of the electric field induced by this emission process depends only on the local magnetic field vector and on the incoming direction of the air shower. We report on measurements of the electric field vector where, in addition to this geomagnetic component, another component has been observed which cannot be described by the geomagnetic emission process. The data provide strong evidence that the other electric field component is polarized radially with respect to the shower axis, in agreement with predictions made by Askaryan who described radio emission from particle showers due to a negative charge-excess in the front of the shower. Our results are compared to calculations which include the radiation mechanism induced by this charge-excess process.
\end{abstract}


\maketitle
\newpage
\newpage
\section{Introduction}\label{sec:introduction}

When high-energy cosmic rays penetrate the atmosphere of the Earth they induce an air shower. The detailed registration of this avalanche of secondary particles is an essential tool to infer properties of the primary cosmic ray, such as its energy, its incoming direction, and its composition. Radio detection of air showers started in the 1960's and the achievements in these days have been presented in reviews by Allan
\cite{Refworks:188} and Fegan \cite{Refworks:189}. In the last decade, there has been renewed interest through the publications of the LOPES \cite{Refworks:101} and CODALEMA \cite{Refworks:91} collaborations. We have deployed and are still extending the Auger Engineering Radio Array (AERA) \cite{Refworks:178,Refworks:180,Refworks:183,vandenBergECR2012,SchroederICRC2013} as an additional tool at the Pierre Auger Observatory to study air showers with an energy larger than $10^{17}$ eV. In combination with the data retrieved from the surface-based particle detectors \cite{Refworks:176} and the fluorescence detectors \cite{Refworks:177} of this observatory, the data from radio detectors can provide additional information on the development of air showers.



An important step in the interpretation of the data obtained with radio-detection methods is the understanding of the emission mechanisms. In the early studies of radio emission from air showers it was conjectured that two emission mechanisms play an important role: the geomagnetic emission mechanism as proposed, amongst others, by Kahn and Lerche \cite{Refworks:190} and the charge-excess mechanism as proposed by Askaryan \cite{Refworks:191}. Essential for the geomagnetic effect is the induction of a transverse electric current in the shower front by the geomagnetic field of the Earth while the charge excess in the shower front is to a large extent due to the knock-out of fast electrons from the ambient air molecules by high-energy photons in the shower. The magnitude of the induced electric current as well as the induced charge excess is roughly proportional to the number of particles in the shower and thus changing in time. The latter results in the emission of coherent radio waves at sufficiently large wavelengths \cite{Refworks:190, REF12a}. The shower front, where both the induced transverse current and the charge excess reside, moves through the air with nearly the velocity of light. Because air has a refractive index which differs from unity, Cherenkov-like time compression occurs \cite{REF12b, REF12c}, which affects both the radiation induced by the transverse current as well as by the charge excess. The polarization of the emitted radiation differs for current-induced and charge-induced radiation, but its direction for each of these individual components does not depend on the Cherenkov-like time compression caused by the refractive index of air. For this reason we will distinguish in this paper only geomagnetic (current induced) and charge-excess (charge induced) radiation. The contribution of the geomagnetic emission mechanism, described as a time-changing transverse current by Kahn and Lerche \cite{Refworks:190}, has been observed and described in several papers; see e.g. Refs. \cite{Refworks:101,Refworks:74,Refworks:35,Refworks:186,Refworks:221}. Studies on possible contributions of other emission mechanisms from air showers have also been reported \cite{Refworks:208,Refworks:209,Refworks:210,Refworks:211,Prescott1971}. An observation of the charge-excess effect in air showers has been reported by the CODALEMA collaboration \cite{Refworks:198}.


We present the analysis of two data sets obtained with two different setups consisting of radio-detection stations (RDSs) deployed at the Pierre Auger Observatory. The first data set was obtained with a prototype setup \cite{Refworks:67} for AERA; the other one with AERA itself \cite{Refworks:178,Refworks:180,Refworks:183} during its commissioning phase while it consisted of only 24 stations. In addition, we will compare these data sets with results from different types of calculations outlined in Refs. \cite{Refworks:3, Refworks:11, Refworks:196, Refworks:220, Refworks:212, Refworks:213, Refworks:223}.

This paper is organized as follows. We discuss in Section \ref{sec:detectionsystem} the experimental equipment used to collect our data. In Section \ref{sec:dataanalysis} we present the data analysis techniques and the cuts that we applied on the data, whilst in Section \ref{sec:comparison} we compare our data with calculations. In Section \ref{sec:conclusion} we discuss the results and we present our conclusions. For clarity, Sections \ref{sec:dataanalysis} and  \ref{sec:comparison} contain only those figures which are based on the analysis for the data obtained with AERA during the commissioning of its first 24 stations; the results of the prototype are shown in Appendix C. We mention that analyses of parts of the data have been presented elsewhere \cite{Refworks:186,Refworks:199,Refworks:214,TimHuegeICRC2013}.

\section{Detection systems}\label{sec:detectionsystem}

\subsection{Baseline detector system}

\begin{figure}[h!t]
	\centering
    \includegraphics[width=0.7\textwidth]{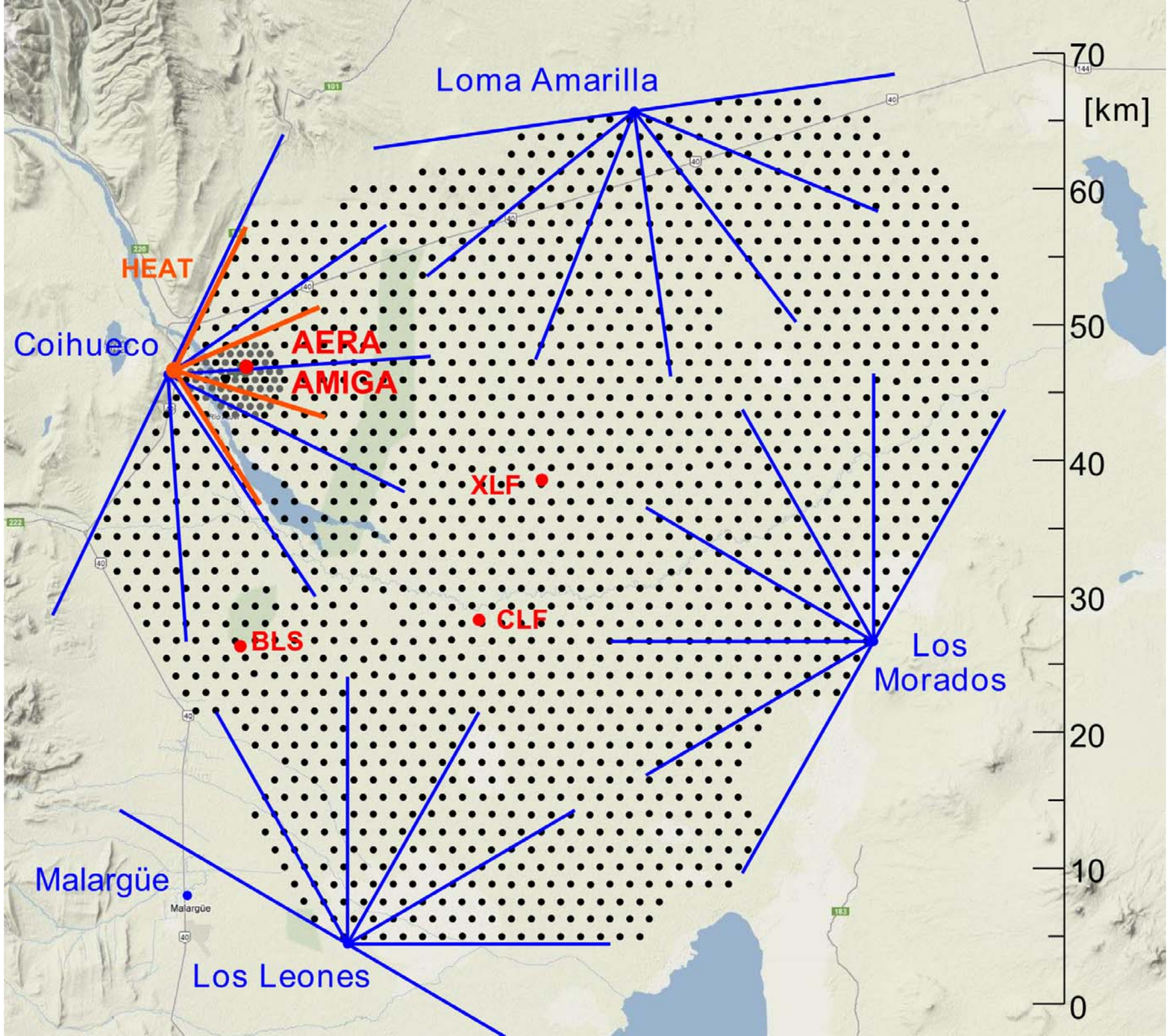} \caption{The detector systems of the Pierre Auger Observatory; the black dots denote the 1660 detector stations of the surface detector (SD), while the buildings containing the telescopes of the fluorescence detector (FD) are located at the edge of the array. The prototype of AERA was located near the Balloon Launching Station (BLS) of the observatory; AERA itself is located in front of the telescope buildings at Coihueco.}
	\label{fig:auger}
\end{figure}

The detection system used at the Pierre Auger Observatory consists of two baseline detectors: the surface detector (SD) and the fluorescence detector (FD), described in detail in Refs. \cite{Refworks:176} and \cite{Refworks:177}, respectively.
The SD is an array consisting of 1660 water-Cherenkov detectors arranged in equilateral triangles with sides of 1.5~km. An infill for the SD, called AMIGA \cite{AMIGAICRC2013}, has been deployed with 750~m spacing between the stations. A schematic diagram of the observatory is shown in Fig. \ref{fig:auger}. In the present study only the SD was used to determine the parameters of the air showers.

\subsection{Radio-detection systems}\label{sec:radiodetector}

\begin{figure}[h!t]
	\centering
	\subfigure{\includegraphics[width=0.45\textwidth, viewport=50 0 600 550, clip]{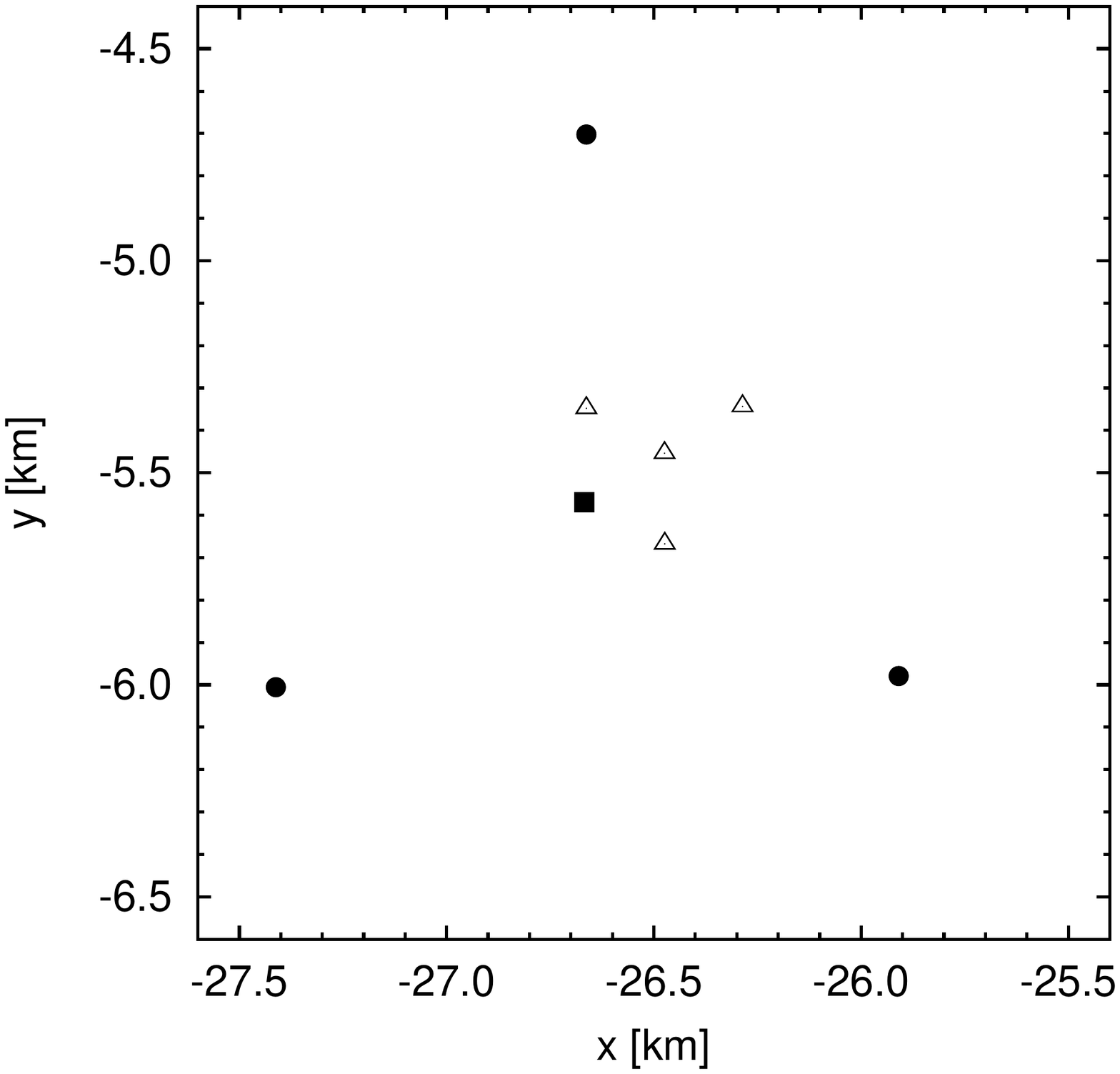}}
    \subfigure{\includegraphics[width=0.45\textwidth, viewport=50 0 600 550, clip]{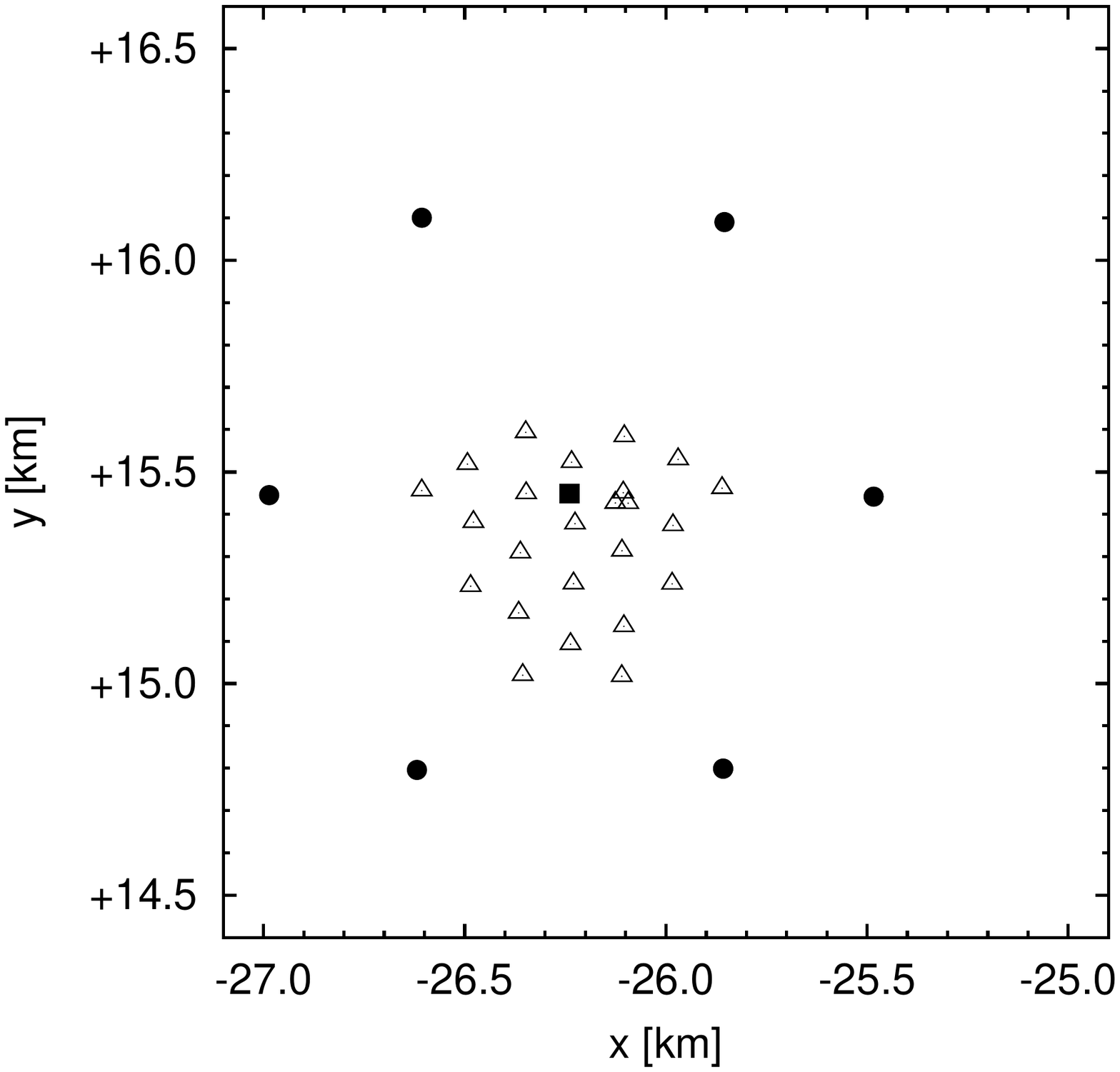}}
	\caption{An aerial view of the radio-detection systems (open triangles) in the SD. Stations of the SD are denoted by filled markers, where the SD stations nearest to the radio-detection systems are denoted with filled squares, for the prototype setup (left panel) and for AERA24 (right panel). The coordinates are measured with respect to the center of the SD.}
	\label{fig:setups}
\end{figure}

The prototype for AERA used in the present study consisted of four RDSs. Each RDS had a dual-polarized logarithmic periodic dipole antenna (LPDA) optimized for receiving radio signals in a frequency band centered at 56~MHz and with a bandwidth of about 50~MHz. These antennas were aligned such that one polarization direction was pointing along the geomagnetic north-south (NS) direction with an accuracy of $0.6^\circ$, while the other polarization direction was pointing to the east-west (EW) direction. For each polarization direction, NS and EW, we used analog electronics to amplify the signals and to suppress strong lines in the HF band below 25~MHz and in the FM-broadcast band above 90~MHz. A 12 bit digitizer running at a sampling frequency of 200 MHz was used for the analog-to-digital conversion of the signals. This electronic system was completed with a GPS system, a trigger system, and a data-acquisition system. The trigger for the station readout was made using a scintillator detector connected to the same digitizer as was used for the digitization of the radio signals. The data from all RDSs were stored on disks and afterwards compared with those from the SD.
To ensure that the events, which have been registered with the RDSs, were indeed induced by air showers, a coincidence between the data from AERA prototype and from the SD in time and in location was required  \cite{Refworks:67}. An additional SD station was deployed near the AERA prototype setup to reduce the energy threshold of the SD; see the left panel of Fig. \ref{fig:setups}. Further details of the AERA prototype stations can be found in Ref. \cite{Refworks:67}.

AERA is sited at the AMIGA infill \cite{AMIGAICRC2013} of the observatory and consists presently of 124 stations \cite{SchroederICRC2013}. The deployment of AERA began in 2010 and physics data-taking started in March 2011. In the data-taking period presented in this work, AERA consisted of 24 RDSs arranged on a triangular grid with a station-to-station spacing of 144~m; see the right panel of Fig. \ref{fig:setups}. For the present discussion, we will denote this stage as AERA24. The characteristics of AERA24 are very similar to its prototype. A comparison of their features is presented in Table \ref{tab:comparison}; see also Ref. \cite{Refworks:202} for further details. One of the main differences between AERA24 and its prototype is that the AERA stations can trigger on the signals received from the antennas whereas the prototype used only an external trigger created by a particle detector.  In addition to these event triggers, both systems recorded events which were triggered every 10~s using the time information from the GPS device of the RDS. These events are, therefore, called minimum bias events.

\begin{table}
\caption{\label{tab:comparison}
    Comparison of characteristic features of AERA24 during this data-taking period and its prototype.}
\begin{ruledtabular}
\begin{tabular}{llll}
                                        &AERA24           &AERA prototype            &\\
    \hline
    antenna type                        &LPDA           &LPDA                       &\\
    number of polarization directions   &2              &2                          &\\
    -3 dB antenna bandwidth             &29 - 83        &32 - 84                    &MHz\\
    gain LNA                            &20             &22                         &dB   \\
    -3 dB pass filter bandwidth         &30 - 78        &25 - 70                    &MHz \\
    gain main amplifier                 &19             &31                         &dB \\
    sampling rate digitizer             &200            &200                        &MHz \\
    digitizer conversion                &12             &12                         &bits \\
    trigger                             &EW polarization&particle                   &\\
    RDS station-to-station spacing      &144            &216                        &m\\
    SD infill spacing                   &750            &866                        &m\\
\end{tabular}
\end{ruledtabular}
\end{table}

\section{Event selection and data analysis}\label{sec:dataanalysis}

The data from the SD and the radio-detection systems were collected and analyzed independently from each other as will be described in sections \ref{sec:sdanalysis} and \ref{sec:rdanalysis}, respectively. Using timing information from both detection systems (see section \ref{sec:coincidences}) an off-line analysis was performed combining the data from both detectors.

\subsection{Conventions} \label{sec:conventions}

For the present analysis we use a spherical coordinate system with the polar angle $\theta$ and the azimuth angle $\phi$, where we define $\theta=0^\circ$ as the zenith direction and where $\phi=0^\circ$ denotes the geographic east direction; $\phi$ increases while moving in the counter-clockwise direction. We determine the incoming direction ($\theta_a$ and $\phi_a$) of the air shower by analysis of the SD data. For the relevant period of data taking, which started in May 2010 and ended in June 2011, the strength and direction of the local magnetic field vector on Earth at the location of the Pierre Auger Observatory was 24~$\mu$T and its direction was pointing to $(\theta_b,\phi_b) = (54.4^\circ,87.3^\circ)$ \cite{Refworks:204}.

\begin{figure}[t]
    \begin{centering}
    \includegraphics[width=0.8\textwidth]{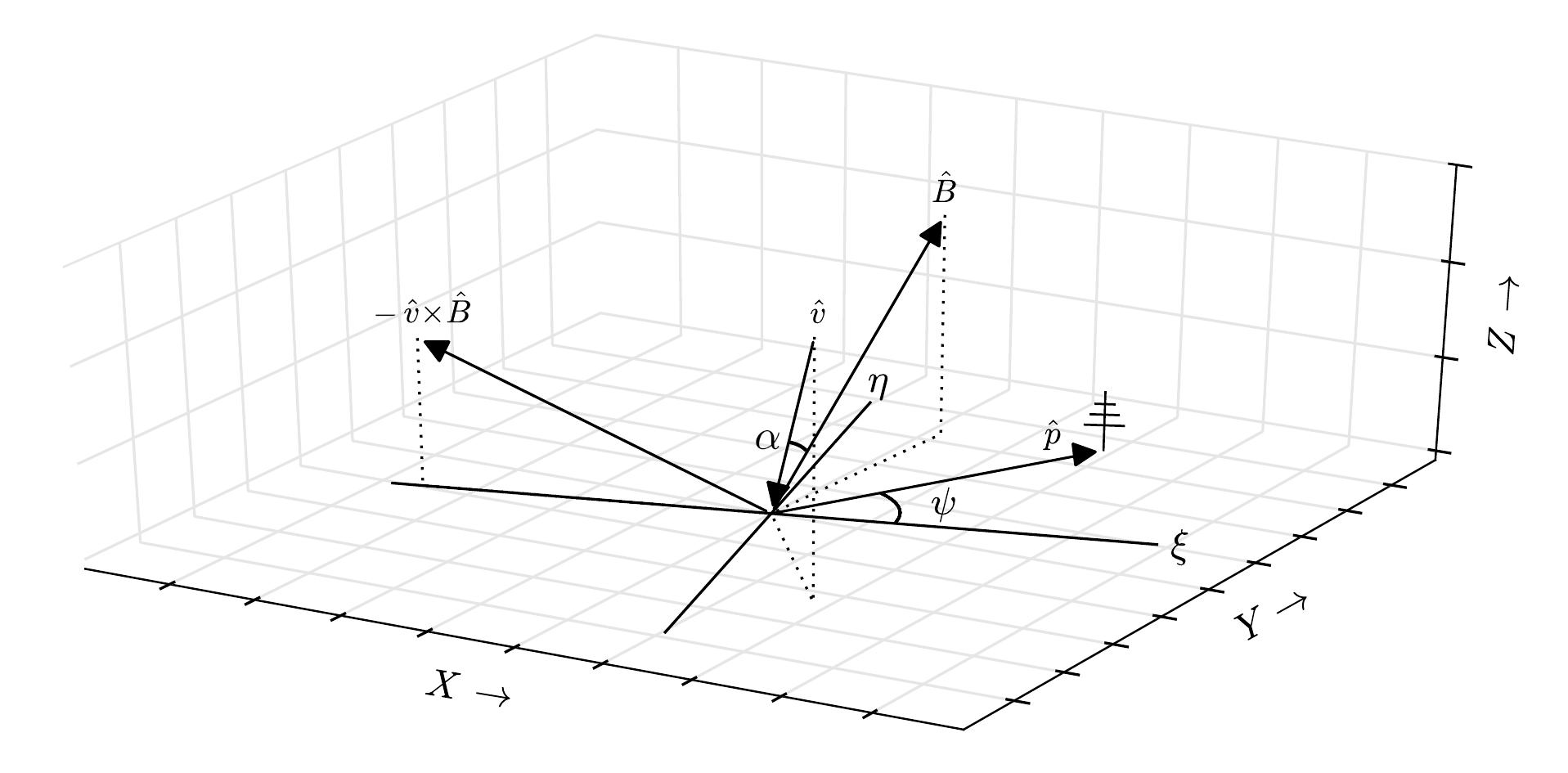}
    \caption{\label{fig:Observer-angle-dependence1} Direction of the incoming shower, denoted as $\hat{v}$, with respect to the position of RDS which is symbolically indicated by an antenna. The direction of the magnetic field vector is denoted by $\hat{B}$ and the direction $\xi$ is defined by the projection of the vector $\hat{v} \times \hat{B}$ onto the ground plane. The direction $\eta$ is perpendicular to $\xi$ and is also in the ground plane. The angle between the shower axis and the geomagnetic field direction is denoted as $\alpha$, while $\psi$ is the azimuthal angle between the $\xi$-axis and the direction of the RDS measured at the core position.}
    \end{centering}
\end{figure}

The contribution to the emitted radio signal caused by the charge-excess effect (denoted as $\vec{E}^{\mathrm{A}}$) is not influenced by the geomagnetic field $\vec{B}$. The induced electric field vector of this effect is radial with respect to the shower axis. As explained in Section \ref{sec:radiodetector} the dual-polarized antennas were directed in the NS and EW directions. Therefore, the relative amplitudes of the registered electric field in each of the two arms of an RDS depend on the position of the RDS with respect to the shower axis. The geomagnetic-emission mechanism induces an electric field $\vec{E}^{\mathrm{G}}$ which is pointing along the direction of $(-\vec{v} \times \vec{B})$ where $\vec{v}$ is a vector in the direction of the shower. Thus, for this emission mechanism, the relative contribution of the registered signals in each of the two arms does not change as a function of the position of the RDS. For this reason, it is convenient to introduce a rotated coordinate system $(\xi, \eta)$ in the ground plane such that the $\xi$ direction is the projection of the vector $(-\hat{v} \times \hat{B})$ onto the shower plane and $\eta$ is orthonormal to $\xi$; see Fig. \ref{fig:Observer-angle-dependence1}. The angle between the incoming direction of the shower and the geomagnetic field vector is denoted as $\alpha$.

For our polarization analysis, we consider a total electric field as the vectorial sum of the geomagnetic and of the charge-excess emission processes:
\begin{equation}
\vec{E}(t)\ =\ \vec{E}^{\mathrm{G}}(t)+\vec{E}^{\mathrm{A}}(t),
\label{eq:evector}
\end{equation}
where $t$ describes the time dependence of the radiation received at the location of an RDS.

\subsection{Data pre-processing and event selection for the SD}\label{sec:sdanalysis}

The incoming directions and core positions of air showers were determined from the recorded SD data. A detailed description of the trigger conditions for the SD array with its grid spacing of 1500~m is presented in Ref. \cite{Refworks:206}. However, as mentioned before, additional SD stations were installed in the neighborhood of the RDSs as infills of the standard SD cell. Because of these additional surface detectors, we used slightly different constraints as compared to the cuts used for the analysis of events registered by the regular SD array only. These additional constraints are a limit on the zenith angle ($\theta<40^\circ$ for the prototype and $\theta<55^\circ$ for AERA24). Furthermore, in the case of events recorded near the prototype, the analysis was based on only those events where the infill SD station near this setup yielded the largest signal strength (i.e., highest particle flux) and where the reconstructed energy by the SD was larger than 0.20 EeV. For the AERA24 events, we required that the distance from the reconstructed shower axis to the infill station was less than 2500~m or that the event contained at least one of the SD stations shown in the right panel of Fig. \ref{fig:setups}. The estimated errors on the incoming direction and on the determination of the core position depend on the energy of the registered air showers. These errors are smaller at higher energies. Typical directional errors in our studies range from $0.5^\circ$ at 1~EeV to $1^\circ$ at 0.1~EeV. The uncertainty in the determination of the core positions also reduces for higher energies and lies around 60~m at 0.1~EeV and around 20~m at 1.0~EeV.

\subsection{Event selection and data analysis for the RDSs}\label{sec:rdanalysis}

\begin{figure}[h!t]
    \centering
        \subfigure{\includegraphics[width=0.45\textwidth, viewport = 0 00 400 300, clip]{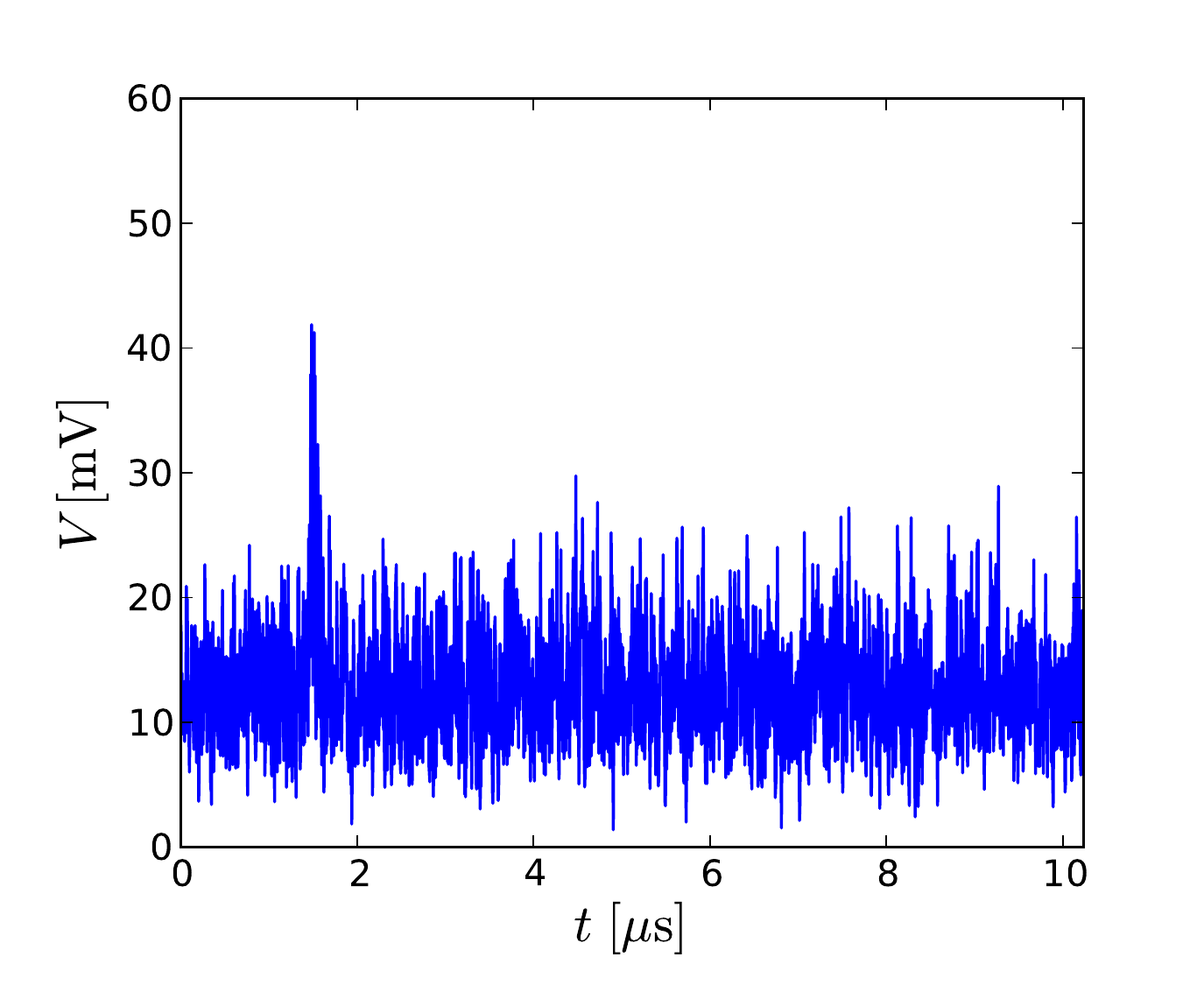}} \\
        \subfigure{\includegraphics[width=0.45\textwidth, viewport = 0 00 400 300, clip]{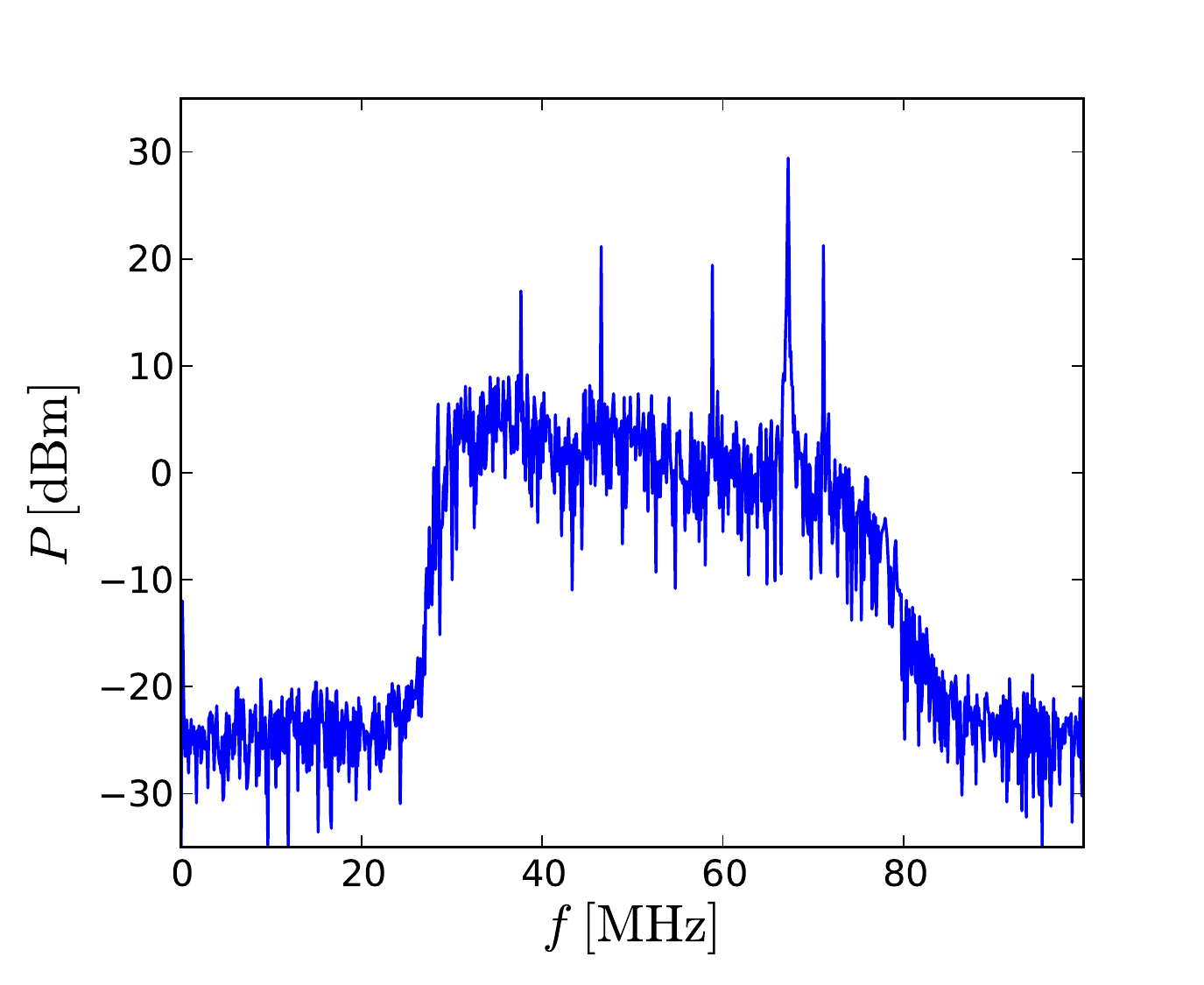}} \\
        \subfigure{\includegraphics[width=0.45\textwidth, viewport = 0 00 400 300, clip]{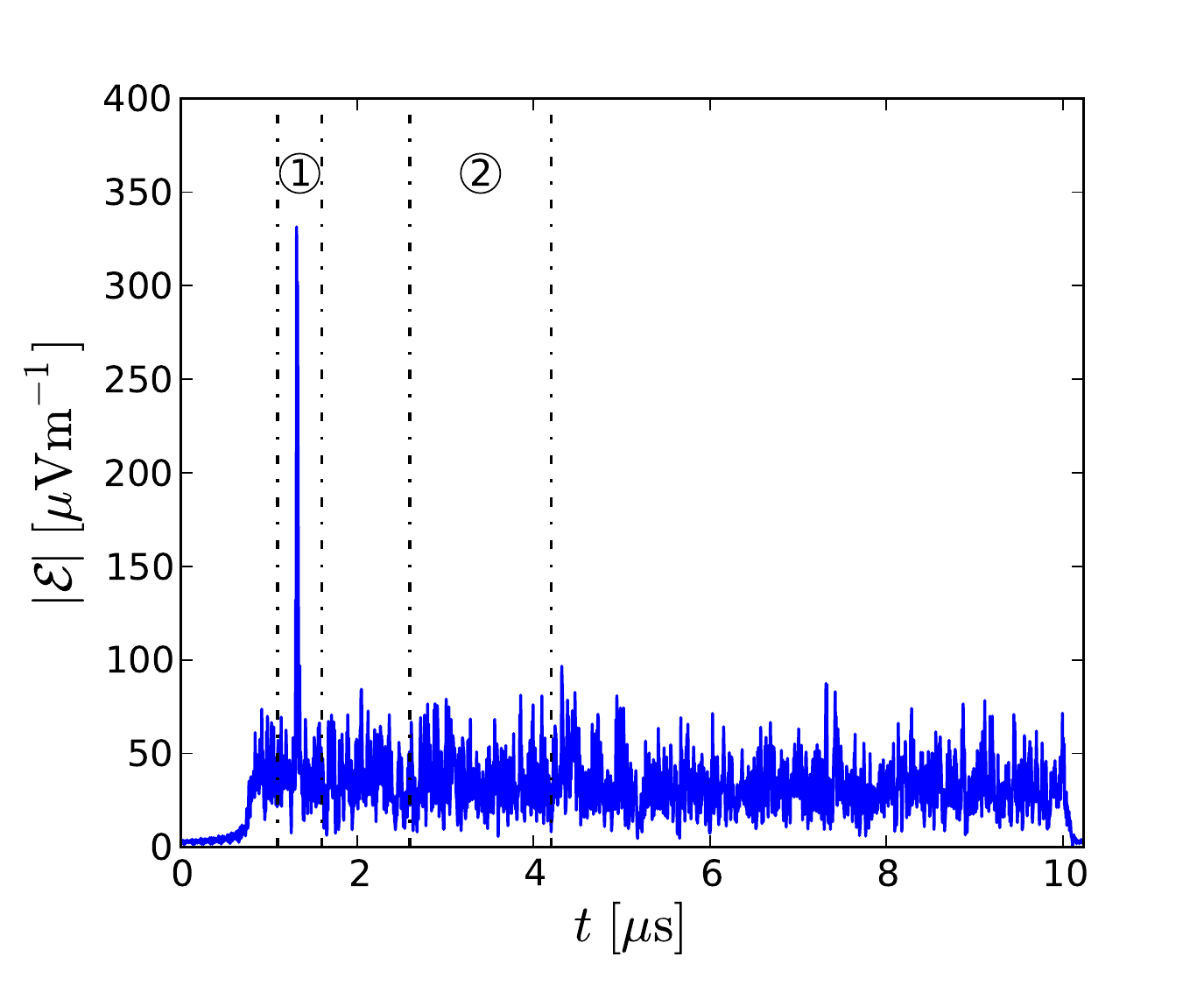}}
        \caption{An example of a radio signal in various stages of the analysis. Upper panel: the square root of the quadratic sum for the signal envelopes of both polarization directions. Middle panel: the power distribution of this signal in the frequency domain. Lower panel: the analytic signal for the electric field $\vec{\mathcal{E}}$ (see Eq. (\ref{eq:hilbertE})) reconstructed from the two time traces and from the incoming direction of the shower. The signal was cleaned from narrow-band transmissions using the linear predictor algorithm. The signal (noise) region used for this algorithm is denoted by 1 (2) and has a width of 125 ns (1600 ns).}
	\label{fig:traces}
\end{figure}

The data from the RDSs were used to determine the polarization of the electric field induced by air showers. Here we take advantage of the fact that the LPDAs were designed as dual-polarized antennas. The data for each of these two polarization directions were stored as time traces with 2000 samples (thus with a length of 10~$\mu s$). We use the Hilbert transformation, which is a standard technique for bandpass-limited signals \cite{Refworks:227}, to calculate the envelope of the time trace. An example of such a trace is shown in the upper panel of Fig. \ref{fig:traces}, which was recorded for an air shower with parameters: $\theta_a = (30.0\pm 0.5)^\circ$, $\phi_a = (219\pm 2)^\circ$ and $E = (0.19 \pm 0.02)$~EeV near the AERA24 site.

We took several measures to ensure good data quality for the received signals in the RDSs. Despite of the bandpass filters (see Table \ref{tab:comparison}) a few narrow-band transmitters contaminated the registered signals. The effect of the suppression of the frequency regions outside the passband and the remaining contributions from the narrow-band transmitters within the passband are displayed in the middle panel of Fig. \ref{fig:traces}, which shows the Fourier transform of the time trace shown in the upper panel. These narrow-band transmitters were removed by applying two different digital methods. The first method operates in the time-domain and involves a linear predictor algorithm based on the time-delayed forecasted behavior of 128 consecutive time samples \cite{Refworks:218}. The second method involves a Fourier- and inverse-Fourier-transform algorithm, where in the frequency domain the power of the narrow-band transmitters was set to zero (see, e.g., Ref. \cite{Refworks:101}).

To determine the total electric field vector, we used the simulated antenna gain pattern \cite{Refworks:201} and the incoming direction of the air showers as determined with the SD analysis. This technique is described in detail in Refs. \cite{Refworks:32,Refworks:179}. In the analysis of the radio signals we used the analytic signal, which is a complex representation of the electric field vector $\vec{\mathcal{E}_{j}}$, where $j$ runs over the sample number in the time trace. This complex vector was constructed from the electric field itself using the Hilbert transformation $\mathcal{H}$:
\begin{equation}
    \vec{\mathcal{E}_{j}} = \vec{E_{j}} + i \mathcal{H}(\vec{E_{j}})
    \label{eq:hilbertE}
\end{equation}
In the lower panel of Fig. \ref{fig:traces} the analytic signal of the electric field for this particular event is displayed after removing the narrow-band transmitters from the signal.

The data obtained with the minimum bias triggers were analyzed to check the gain for the different polarization directions. For this comparison we used the nearly daily variation of the signal strength in each RDS (for the two polarization directions) caused by the rising and setting of strong sources in the galactic plane; see e.g. Ref. \cite{Refworks:67}. In the present analysis only data from those RDSs were selected where the difference between the relative gain for their two polarization directions was less than 5\%.
Furthermore, it is well-known \cite{Refworks:85,Refworks:225,AUGERJINST} that thunderstorm conditions may cause a substantial change of the radio signal strength from air showers as compared to the signal strength obtained under fair-weather conditions. The atmospheric monitoring systems of the observatory, located at the BLS and at AERA (see Fig. \ref{fig:auger}) register the vertical electric field strength at a height of about 4~m. Characteristic changes in the static vertical electric field strength are indicative for thunderstorm conditions and air-shower events collected during such conditions have been ignored in the present analysis.

\subsection{Coincidences between the SD and the RDSs}\label{sec:coincidences}

\begin{figure}[h!t]
	\centering
	\includegraphics[width=0.6\textwidth]{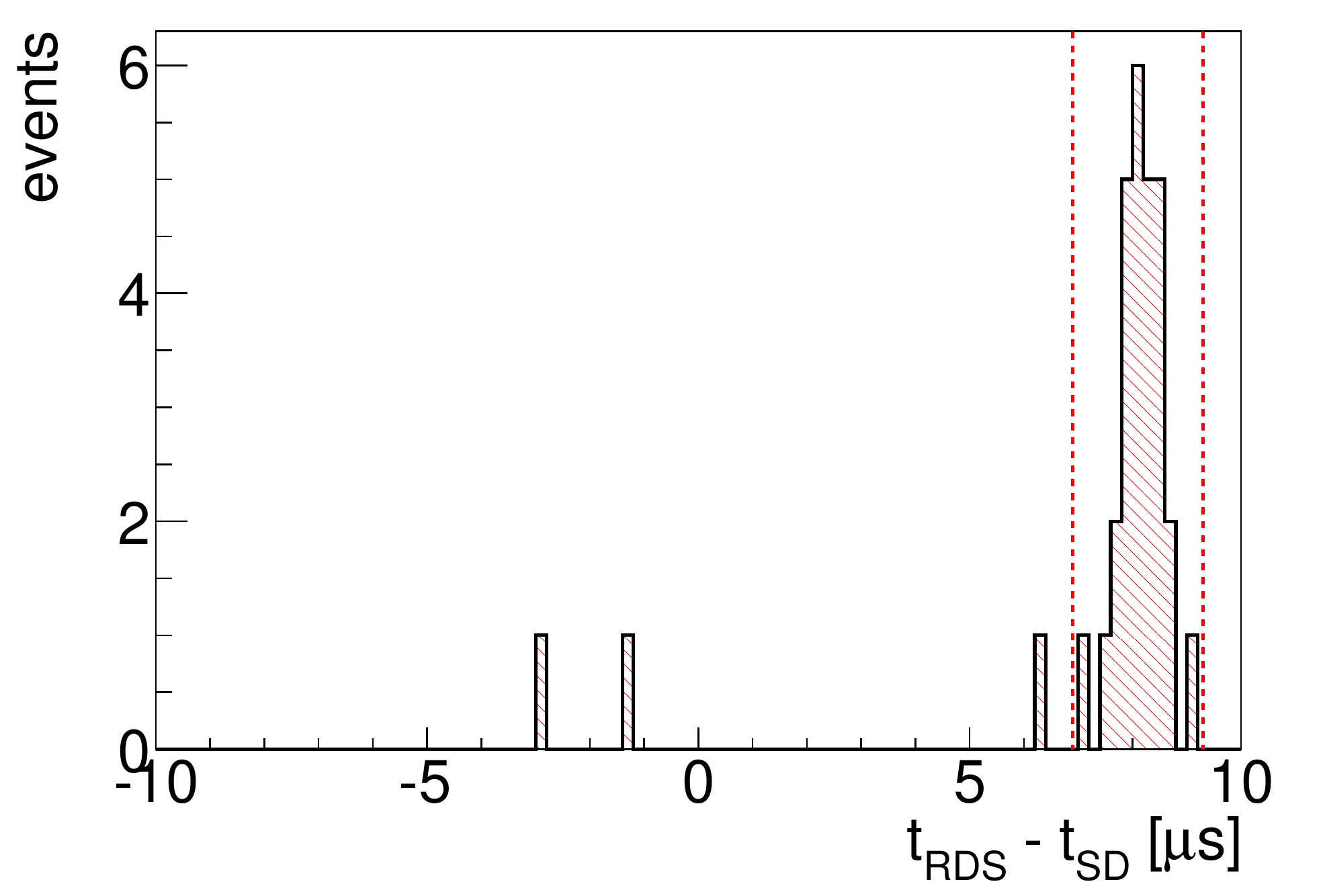}
    \caption{Difference between the reconstructed arrival time of the air-shower events recorded by the SD and the RDSs of AERA24.}
	\label{fig:reltiming}
\end{figure}

The data streams for the SD and RDSs were checked for coincidences in time and in location. For the SD events we used the reconstructed time at which the shower hits the ground at the core position. For the timing of an event registered by one or more of the RDSs we used the earliest time stamp obtained from the triggered stations. We required that the relative difference in the timing between the SD events and the events registered by the radio detector is smaller than 10~$\mu$s. The distribution of the relative time difference of the coincident events registered with AREA is shown in Fig. \ref{fig:reltiming}. The shift of about $8~\mu$s between the SD and RDS timing is caused by the different trigger definitions used for each of the two different detection systems: SD and RDS. This figure clearly displays the prompt coincidence peak and some random events. The events selected for further analysis are within the indicated window in this figure.

\begin{figure}[h!t]
	\centering
	\subfigure{\includegraphics[width=0.70\textwidth, viewport=20 0 550 400, clip]{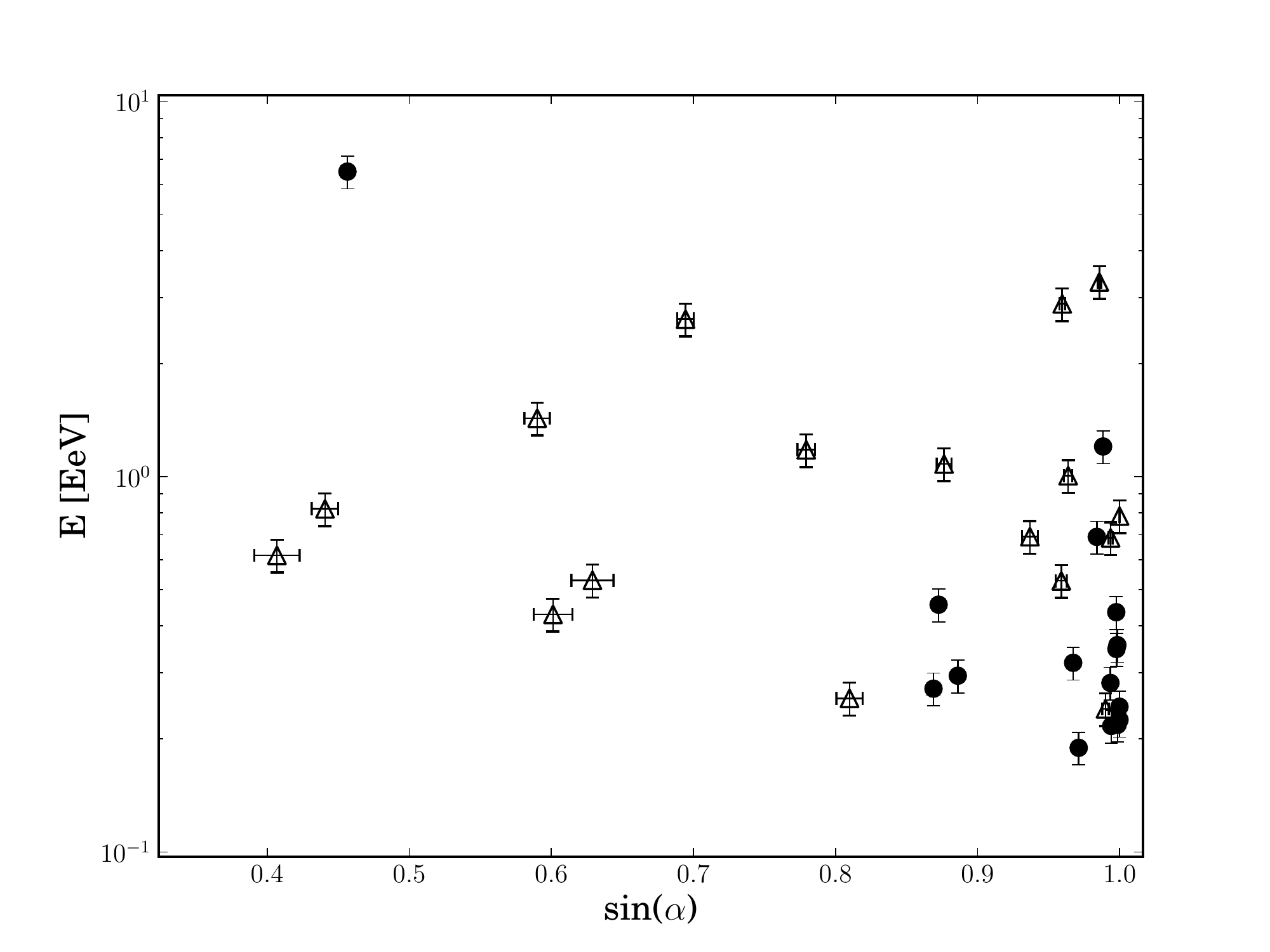}} \\
    \subfigure{\includegraphics[width=0.70\textwidth, viewport=20 0 550 400, clip]{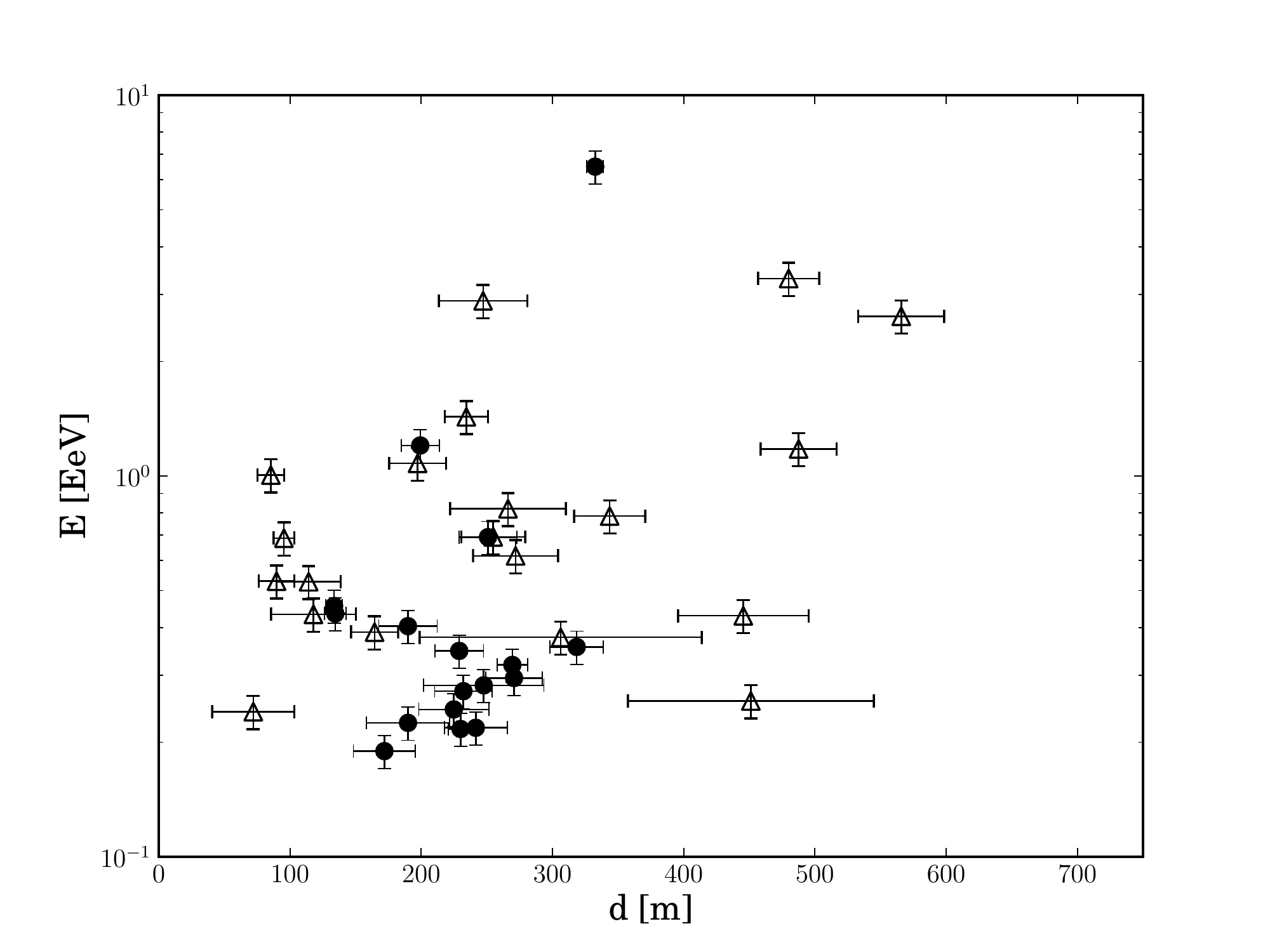}}
	\caption{Scatter plot of shower parameters for coincident events used in the analysis; the filled circles (open triangles) are data for AERA24 (prototype). Upper panel: the shower energy $E$ reconstructed from the SD information versus the space angle $\alpha$ between the magnetic field vector and the shower axis. Lower panel: the reconstructed energy $E$ versus the distance $d$ between the shower axis and the SD station located closest to the radio-detection systems (see Fig. \ref{fig:setups}).}
	\label{fig:EversusA}
\end{figure}

Our analysis is based on 37 air-shower events, 17 registered with AERA24, the other 20 registered with the prototype. We note here, that each event can produce several data points in our analysis. The distance $d$ between the shower axis and the SD station closest to the center of the radio setup, the angle $\alpha$ between the magnetic field vector and the shower direction, and the shower energy $E$ are relevant parameters for the RDS triggers discussed in Section \ref{sec:radiodetector}. The upper (lower) panel of Fig. \ref{fig:EversusA} displays a scatter plot of these coincident events in the $(E, \alpha)$ and $(E,d)$ parameter space.

\subsection{Deviation from geomagnetic polarization as a function of the observation angle}\label{sec:r_analysis}

For each shower and each RDS, we used the SD time and the $(x,y)$ coordinates of each RDS to define a region of interest with a width of 500~ns in the recorded RDS time traces. In this region, indicated in the bottom panel of Fig. \ref{fig:traces} as region 1, we expect radio pulses from recorded air-shower events. Because of time jitter the precise location of the radio signal itself was determined using a small sliding window with a width of $125~\mathrm{ns}$, i.e. 25 time samples. As a first step, we removed the narrow-band noise using the method based on a linear prediction, described in Section \ref{sec:radiodetector}. Then, within the region of interest, the total amplitude in the sliding window was computed by averaging the squared sum of the three electric field components and taking the square root of this summed quadratic strength. The maximum strength obtained by the sliding window was then chosen to be the signal $S$. Thus we have
\begin{equation}\label{eq:S}
S = \frac{1}{25} \left(\sum_{j=1}^{25}\vec{\mathcal{E}}_{j+k} \cdot \vec{\mathcal{E}^{*}}_{j+k}\right)^{1/2},
\end{equation}
where the left edge of this sliding window has sample identifier $k$. For every measured trace, $k$ was chosen such that $S$ reaches a maximum value. The noise level $N$ was determined from region 2 shown in the bottom panel Fig. \ref{fig:traces}. This region has a width of $1600~\mathrm{ns}$ (320 samples).
\begin{equation}\label{eq:N}
N = \frac{1}{320} \left({\sum_{m=m_0}^{m_1}\vec{\mathcal{E}}_{m} \cdot \vec{\mathcal{E}^{*}}_{m}}\right)^{1/2},
\end{equation}
where $m_0$ and $m_1 = m_0 + 319$ are respectively the start and stop samples of this noise region. For the AERA24 data shown in the bottom panel of Fig. \ref{fig:traces} the value of $m_0 = 520$ ($1600~\mathrm{ns}$).

\begin{figure}[t]
    \centering
    \includegraphics[height=0.7\textwidth]{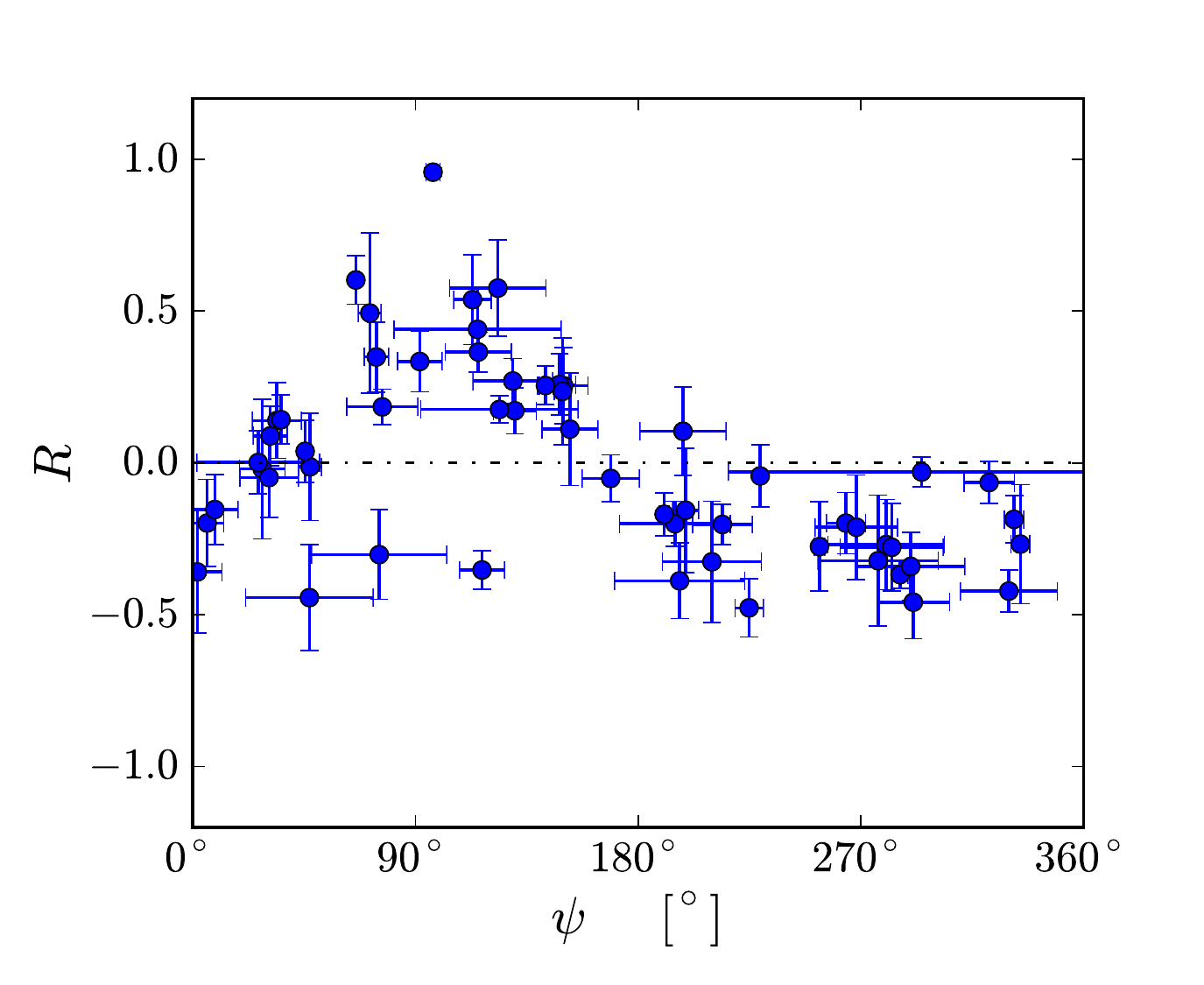}
    \caption{The calculated value of $R$ (see Eq. (\ref{eq:observable})) and its uncertainty for the AERA24 data set as a function of the observation angle $\psi$. The dashed line denotes $R=0$.}
    \label{fig:Observer-angle-dependenceAERA}
\end{figure}

To identify the two mechanisms of radio emission under discussion, we take advantage of the different polarization directions that are expected in each case (see Section \ref{sec:conventions}) where we introduced the coordinate system $(\xi, \eta)$ depicted in Fig. \ref{fig:Observer-angle-dependence1}. In the rotated coordinate system the resulting strength of the electric field in the ground plane is given as: $\mathcal{E}_{\xi}$ and $\mathcal{E}_{\eta}$. We define the quantity $R$ as:
\begin{equation}\label{eq:observable}
R(\psi) \equiv \frac{2\sum_{j=1}^{25}\mathrm{Re}(\mathcal{E}_{j+k, \xi}\; \; \mathcal{E}^{*}_{j+k, \eta})}{\sum_{j=1}^{25}(|\mathcal{E}_{j+k, \xi}|^{2}+|\mathcal{E}_{j+k, \eta}|^{2})},
\end{equation}
where we use the observation angle $\psi$, which is the azimuthal angle at the shower core between the position of the RDS and the direction of the $\xi$-axis. Since $\mathcal{E}_{\eta}$ has no component in the case of pure geomagnetic emission it is clear from Eq. (\ref{eq:observable}) that any measured value of $R \neq 0$ indicates a component different from geomagnetic emission. The measured value of $R$ incurs a bias in the presence of noise, which was taken care of using the procedure explained in Appendix \ref{app:A}. To use signals with sufficient quality the following signal-to-noise cut was used:
\begin{equation}\label{eq:StoN}
S/N > 2
\end{equation}
where $S$ and $N$ are defined by Eqs. (\ref{eq:S}) and (\ref{eq:N}), respectively. The uncertainties in $R$ were obtained by adding noise from the defined noise region to the signal, such that a set of varied signals was obtained:
\begin{equation}\label{eq:Efornoise}
\vec{\mathcal{E}}'_{i+k} = \vec{\mathcal{E}}_{i+k} + \vec{\mathcal{E}}_{i+m}
\end{equation}
for $i$ running from 1 to 25 for each value of $m$ in the noise region from $m_0$ to $m_0+294$. We recall, that the index $k$ was chosen such that $S$ reaches its maximum value (see Eq. (\ref{eq:S})).
Similar to the determination of the value $R$ using Eq. (\ref{eq:observable}), a set of values $R'$ was generated using the values $\vec{\mathcal{E}}'_{i+k}$. From their probability density function the variance and the spread $\sigma_R$ for $R$ were determined. The uncertainty $\sigma_\psi$ in the observation angle was determined from the SD data and from the location of the RDS relevant for the data point plotted. The values of $R$ and their uncertainties are displayed in Fig. \ref{fig:Observer-angle-dependenceAERA} as a function of the angle $\psi$ for the events recorded by AERA24 which passed all the quality cuts.
This $\psi$-dependence of $R$ reflects predictions made by Refs. \cite{Refworks:220,Refworks:3}. These predictions are based on simulations which account for geomagnetic emission and emission induced by the excess of charge at the shower front. Therefore, Fig. \ref{fig:Observer-angle-dependenceAERA} gives evidence that the emission measured cannot be ascribed to the geomagnetic emission mechanism alone.

\subsection{Direction of the electric field vector}\label{sec:a_analysis}

To quantify the deviation from the geomagnetic radiation as measured with our equipment, we compared measured polarization angles with predictions based on a simple model. This model assumes a contribution, in addition to the geomagnetic process, which has a polarization like the one from the charge-excess emission process. From Eq. (\ref{eq:evector}) we can write the azimuthal polarization angle as:
\begin{align}\label{eq:pre_pol_ang2}
\phi_p &= \tan^{-1}\left(\frac{E^{\mathrm{G}}_y + E^{\mathrm{A}}_y}{E^{\mathrm{G}}_x + E^{\mathrm{A}}_x}\right) \notag \\
    &=\tan^{-1}
    \left(\frac
    {\sin(\phi^{\mathrm{G}}) \sin(\alpha) + a \sin(\phi^{\mathrm{A}})}
    {\cos(\phi^{\mathrm{G}}) \sin(\alpha) + a \cos(\phi^{\mathrm{A}})  }\right).
\end{align}
Here $\phi^{\mathrm{G}}$ is the azimuthal angle of the geomagnetic contribution with respect to the geographic east; similarly, $\phi^{\mathrm{A}}$ gives the azimuthal angle for the charge-excess emission. The subscripts $x$ and $y$ denote the geographic east and north directions, respectively; see Fig. \ref{fig:Observer-angle-dependence1}. From the incoming direction of the air shower and the direction of the geomagnetic field ($-\vec{v} \times \vec{B}$) we obtained $\phi^{\mathrm{G}}$.
Using the zenith angle $\theta_a$ and the azimuthal angle $\phi_a$ of the shower axis as well as the angle $\psi$, we define the azimuthal angle $\phi^{\mathrm{A}}$
as:
\begin{equation}\label{eq:phic}
\phi^{\mathrm{A}} = \tan^{-1}
    \left(
    \frac{
        \sin^2(\theta_a) \cos(\psi-\phi_a) \sin(\phi_a) - \sin(\psi)}{
        \sin^2(\theta_a) \cos(\psi-\phi_a) \cos(\phi_a) - \cos(\psi)
    }
    \right)
\end{equation}
while taking into account the signs of the numerator and the denominator in this equation.
In Eq. (\ref{eq:pre_pol_ang2}) the parameter $a$ gives the relative strength of the electric fields induced by the charge-excess and by the geomagnetic emission processes:
\begin{equation}\label{eq:definitionofa}
a \equiv \sin(\alpha)\frac{|E^{\mathrm{A}}|}{|E^{\mathrm{G}}|}.
\end{equation}

To obtain the azimuthal polarization angle from the observed electric field (see Section \ref{sec:rdanalysis}) we used a formalism based on Stokes parameters, which are often used in radio astronomy; see e.g. Ref. \cite{stokesparameters} for more details. Using Eq. (\ref{eq:hilbertE}), the EW and NS components are presented in a complex form, $\mathcal{{E}}_{j,x} = E_{j,x} + i\tilde{E}_{j,x}$ and $\mathcal{{E}}_{j,y} = E_{j,y} + i\tilde{E}_{j,y}$, where we use the notation: $\tilde{E} = \mathcal{H}(E)$ and where $j$ denotes the sample number (i.e. time sequence). In this representation the time-dependent intensity of the electric field strength is given by:
\begin{equation}
I_j \equiv  E^2_{j,x} + \tilde{E}^2_{j,x} + E^2_{j,y} + \tilde{E}^2_{j,y}
\end{equation}
After removing the contributions from narrow-band transmitters using the noise reduction method based on transformations forth and back to the frequency domain (see Section \ref{sec:radiodetector}), we used the region of interest displayed in the bottom panel of Fig. \ref{fig:traces} to find the signal. Because the recorded pulses induced by air showers were limited in time, the average polarization properties were calculated in a narrow signal window. The position and width of this window were defined as the maximum and the full width at half maximum (FWHM) of the intensity of the signal. In this signal window the two Stokes parameters that represent the linear components were calculated as:
\begin{eqnarray}
Q & = & \frac{1}{n}\sum_{j=1}^n (E^2_{j,x} + \tilde{E}^2_{j,x} - E^2_{j,y} - \tilde{E}^2_{j,y})\\
U & = & \frac{2}{n}\sum_{j=1}^n (E_{j,x} \; E_{j,y} + \tilde{E}_{j,x} \; \tilde{E}_{j,y}),
\end{eqnarray}
with $n$ the number of samples for the FWHM window. The uncertainties on $Q$ and $U$, due to uncorrelated background, are given by:
\begin{eqnarray}
\label{eq:sigmaS1}
\sigma^{2}_{Q} &= &\frac{16}{n^2}\sum_{j=1}^{n}\sum_{k=1}^{n}\Big( E_{j,x}\; E_{k,x} \; \text{Cov}(E_{j,x} \; , \; E_{k,x}) + E_{j,y} \; E_{k,y} \; \text{Cov}(E_{j,y} \; , \; E_{k,y})\Big)\\
\label{eq:sigmaS2}
\sigma^{2}_{U} &= &\frac{16}{n^2}\sum_{j=1}^{n}\sum_{k=1}^{n}\Big( E_{j,y}\; E_{k,y} \; \text{Cov}(E_{j,x} \; , \; E_{k,x}) +  E_{j,x} \; E_{k,x} \; \text{Cov}(E_{j,y} \; , \; E_{k,y})\Big),
\end{eqnarray}
in which Cov$(E_{j,x}\; , \; E_{k,x})$ is the covariance between sample $E_{j,x}$ and $E_{k,x}$. The covariance between samples was estimated in a time window that contains only background. It was checked that the contribution from a cross correlation between the $E_{j,x}$ and $E_{k,y}$ samples can be neglected. From $Q$ and $U$ the polarization angle for each recorded shower and at each RDS was obtained using:
\begin{equation}
\phi_p = \frac{1}{2}\tan^{-1}\left(\frac{U}{Q}\right),
\end{equation}
in which the relative sign of $Q$ and $U$ should be taken into account. The uncertainty on $\phi_p$ is given by
\begin{equation}\label{eq:sigmaphip}
\sigma_{\phi_{p}}= \sqrt{\frac{\big(\sigma_{Q}^2 \; U^2+\sigma_{U}^2 \; Q^2 \big)}{4(U^2 + Q^{ 2})^2}}.
\end{equation}
To assure good data quality for the registered time traces, only signals were considered that pass the following signal to noise cut
\begin{equation}\label{eq:SNRL}
\frac{S}{N} = \sqrt{\frac{Q^2 + U^2}{\sigma_{Q}^2+ \sigma_{U}^2}} > 2
\end{equation}

\begin{figure}[ht!]
   \centering
   \includegraphics[width=.8\textwidth]{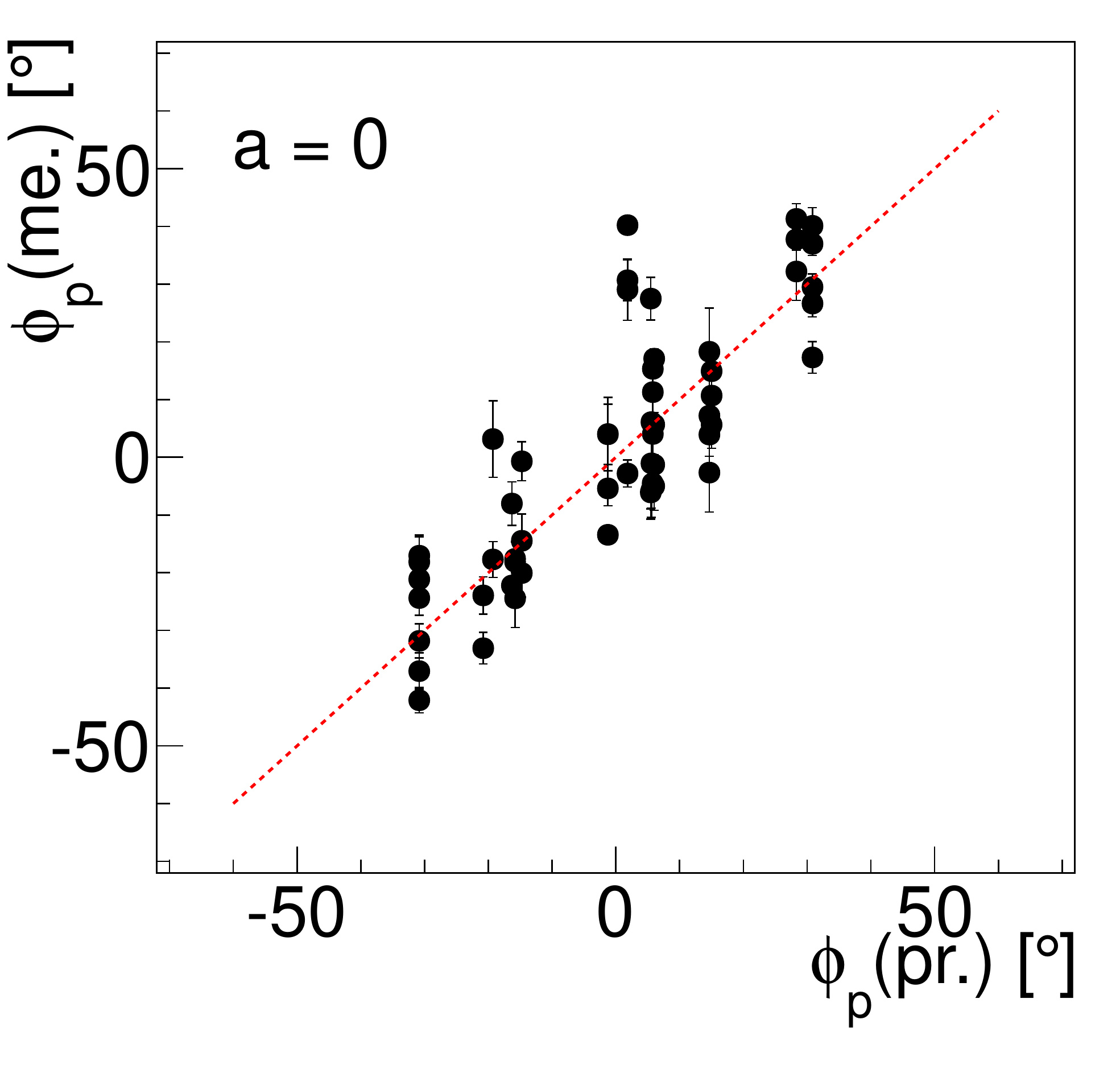}
   \caption{The measured polarization angle versus the predicted polarization angle for the AERA24 data set assuming pure geomagnetic emission: $a = 0$ (see Eqs. (\ref{eq:pre_pol_ang2}) and (\ref{eq:definitionofa})). The dashed line denotes where $\phi_p(\mbox{me.}) = \phi_p(\mbox{pr.})$.}
   \label{fig:polarizationAERA0}
\end{figure}

For the data obtained with AERA24, we compare in Fig.  \ref{fig:polarizationAERA0} the measured polarization angle $\phi_p\mbox{(me.)}$ to the predicted polarization angle $\phi_p\mbox{(pr.)}$ as expected from a pure geomagnetic emission mechanism ($a = 0$). The error bar on the measured value $\phi_p\mbox{(me.)}$ was calculated from Eq. (\ref{eq:sigmaphip}), while the error bar on the predicted value $\phi_p\mbox{(pr.)} = \phi_G$ was obtained from the propagation of the uncertainties on $-\vec{v}$ and $\vec{B}$ in Eq. (\ref{eq:pre_pol_ang2}). Note that this latter uncertainty is always smaller than the size of the marker.

The data displayed in Fig. \ref{fig:polarizationAERA0} show that there is a correlation between the predicted and the measured values of the polarization angle $\phi_p$. The Pearson correlation coefficient for this data set is  $\rho_P = 0.82 ^{+0.06} _{-0.04}$ at 95\% CL. This provides a strong indication that the dominant contribution to the emission for the recorded events was caused by the geomagnetic emission process. As a measure of agreement between the measured and predicted values we calculated the reduced $\chi^2$ value:
\begin{equation}
\frac{\chi^2}{\text{ndf}} = \frac{1}{N}\sum^N\frac{\big(\phi_p \text{(pr.)} - \phi_p \text{(me.)}\big)^2}{\sigma_{\phi_p}^2\text{(pr.)} + \sigma_{\phi_p}^2 \text{(me.)}},
\end{equation}
where the sum runs over all $N$ measurements. For the case where $a=0$ the value of $\chi^2/\text{ndf} = 27$.

The value of $a$ per individual measurement can be determined using Eq. (\ref{eq:pre_pol_ang2}). This equation was used to predict the value of $\phi_p$ by varying the value of $a$ over a wide range. From this scan we obtained a most probable value for $a$ and its 68\% (asymmetric) uncertainty; for details see Appendix \ref{app:tech_a}. For computational reasons we limited ourselves to the range $-1\leq a\leq1$, where $a = -1 (+1)$ corresponds to a radial outwards (inwards) polarized signal that equals to strength of the geomagnetic contribution.

\begin{figure}[ht!]
   \centering
   \includegraphics[width=.6\textwidth]{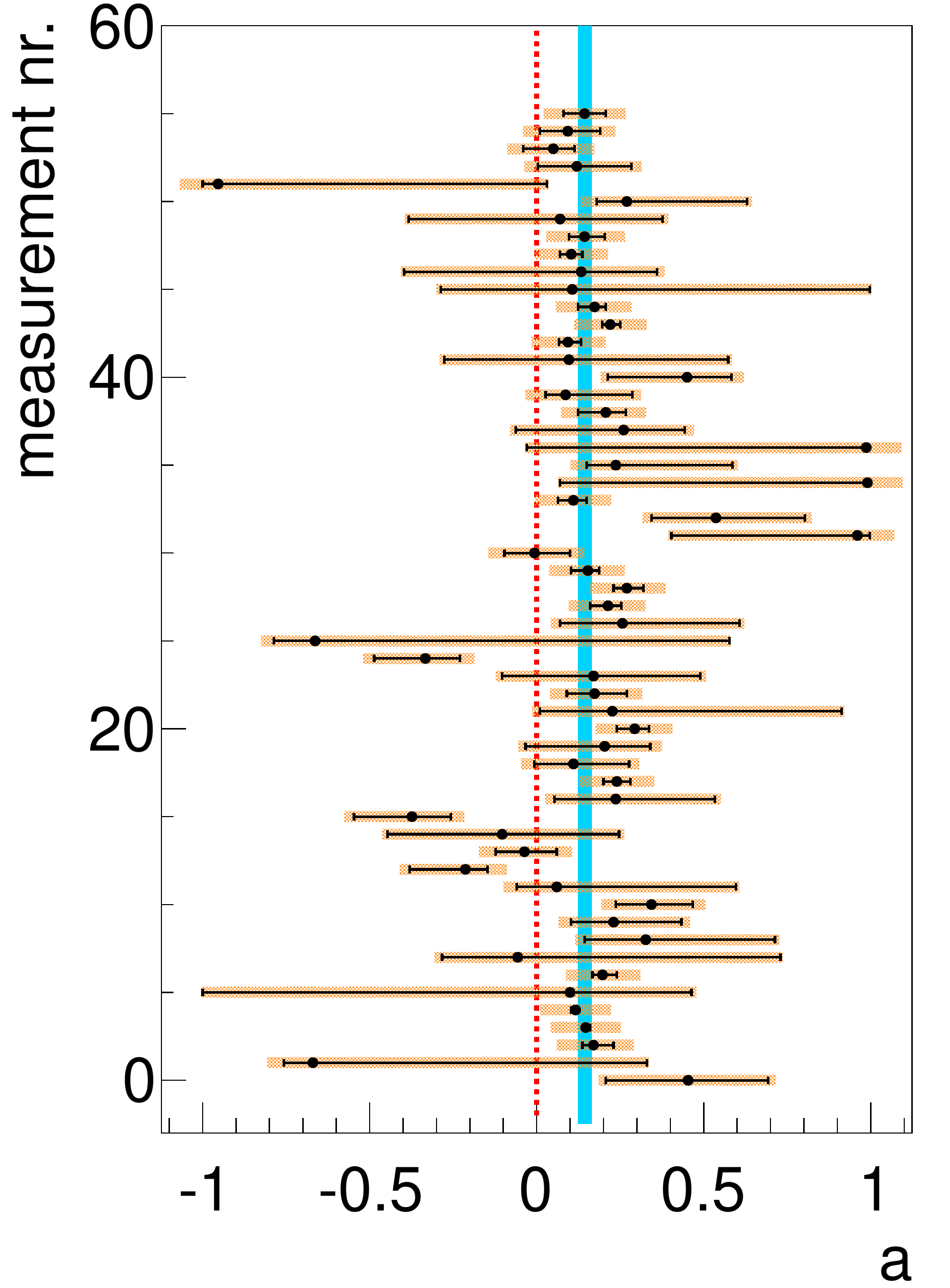}
   \caption{Distribution of most probable values of $a$ (see Eq. (\ref{eq:definitionofa})) and their uncertainties for the AERA24 data set (see Appendix B for details). The 68\% confidence belt around the mean value of $a$ is shown as the solid blue line; the value $a=0$ is indicated with the dotted red line; see text for further details.}
   \label{fig:a-aera}
\end{figure}

In Fig. \ref{fig:a-aera}, the estimated value of $a_i$ and its estimated uncertainty $\sigma_i$ per measurement is shown by the black markers and error bars, respectively. From Fig. \ref{fig:a-aera} it is clear that, given the uncertainties on the measurements, the values of $a_i$ do not arise from a single constant value of $a$. The reason might be that the values of $a$ depend on more parameters, such as on the distance to the shower axis and/or on the zenith angle. From the distribution of $a_i$ values, we estimate the mean value. We do this by taking into account the additional spread in the sample $\Delta$ by requiring that
\begin{equation}
\frac{\chi^2}{\text{ndf}} = \frac{1}{n}\sum^N_i \frac{(a_i - \bar{a})^2}{(\sigma_{a_i}^2 + \Delta^2)} = 1.
\label{eq:dem}
\end{equation}
In which the mean value $\bar{a}$ is calculated using a weighted average, with weights
\begin{equation}
w_i =1/(\sigma_{a_i}^2 + \Delta^2).
\end{equation}
We use for $\sigma_{a_i}$ the upper uncertainty bound when $\bar{a} $ is larger than $a_i$, and the lower uncertainty bound when $\bar{a}$ is smaller the $a_i$. We find that the requirement in Eq. (\ref{eq:dem}) is satisfied at a value $\Delta = 0.10$, and the rescaled uncertainties $\sqrt{(\sigma_{a_i}^2 + \Delta^2)}$ are indicated by the orange boxes around the data points in Fig. \ref{fig:a-aera}. The mean value is estimated to be $\bar{a} = 0.14$, the uncertainty on the mean is estimated from the weights
\begin{equation}
\sigma_{\bar{a}} = \frac{1}{\sqrt{\sum_i^n w_i}}.
\end{equation}
and has a value 0.02.

\begin{figure}[ht!]
   \centering
   \includegraphics[width=.8\textwidth]{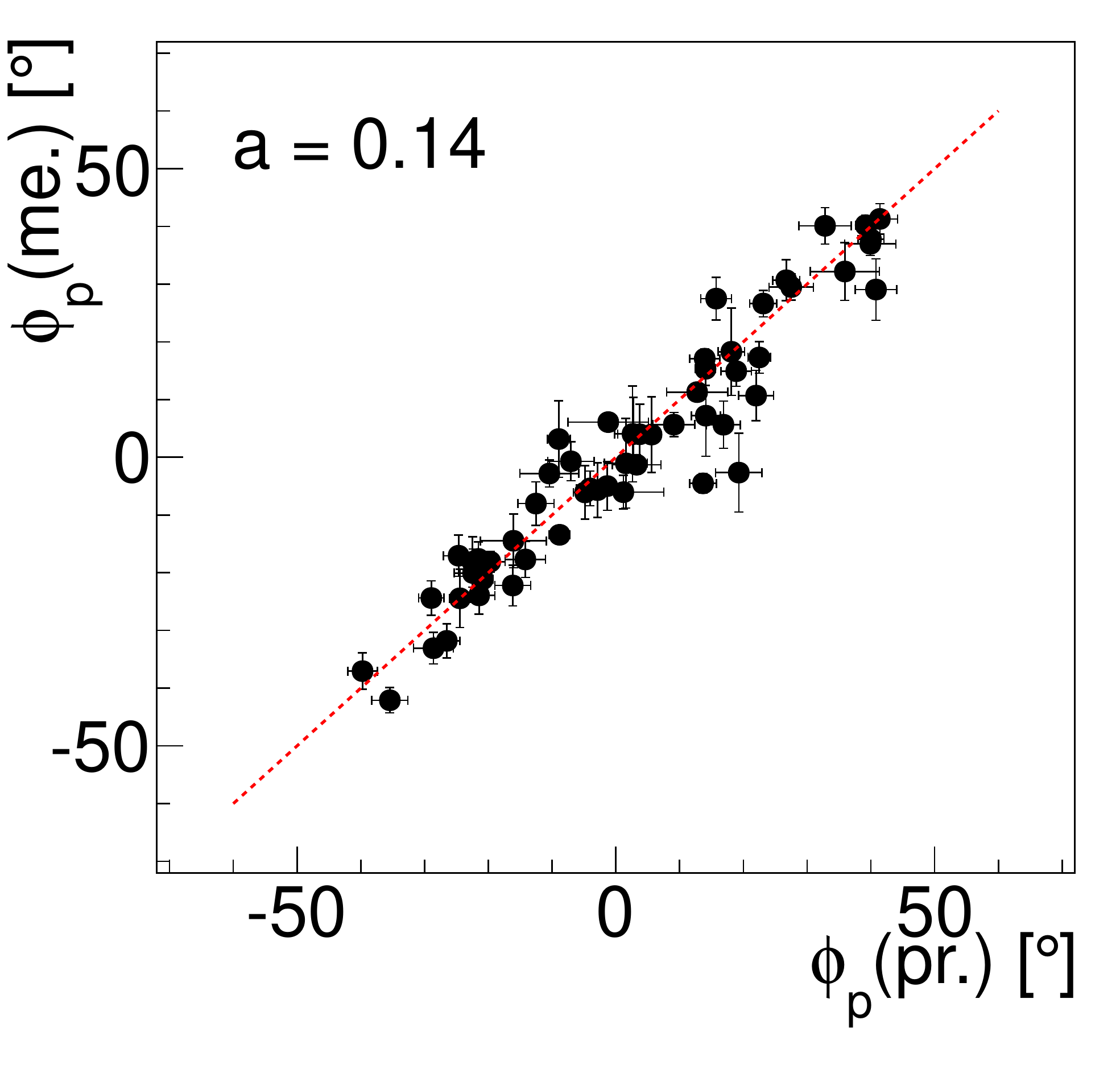}
    \caption{The predicted polarization angle using the combination of the two emission mechanisms with  $a = 0.14\pm0.02$, versus the measured polarization angle for the AERA24 data set; see also the caption to Fig. \protect{\ref{fig:polarizationAERA0}}.}
   \label{fig:polAng-a0123}
\end{figure}

The deduced mean value of $a$ has been used to predict with Eq. (\ref{eq:pre_pol_ang2}) the values of $\phi_p$ and its uncertainty based on the uncertainties in the location and the direction of the shower axis and on the uncertainty in the direction of the geomagnetic field. These predictions are shown in Fig. \ref{fig:polAng-a0123} and compared to the measured polarization angles. In the case where we take $a = 0.14$, the value of the Pearson correlation coefficient is given by $\rho_P = 0.93 ^{+0.04} _{-0.03}$ at 95\% CL. If we compare this number with the value obtained under the assumption, that there is only geomagnetic emission ($a=0$ with $\rho_P = 0.82 ^{+0.06} _{-0.04}$; see Fig. \ref{fig:polarizationAERA0}), we see that the correlation coefficient increases significantly. In addition, the reduced $\chi^2$-value decreases from 27 for $a = 0.0$ to 2.2 for $a = 0.14$.

This deduced contribution for a radial component with a strength of $(14 \pm 2)$\% compared to the component induced by the geomagnetic-emission process is, within the uncertainties, in perfect agreement with the old data published in Refs. \cite{Refworks:209,Prescott1971}. They quote values of $(15 \pm 5)$\% and $(14 \pm 6)$\% for a radio-detection setup located in British Colombia and operated at 22 MHz.

\clearpage
\subsection{Summary of experimental results}\label{sec:expsummary}

The results presented in the previous sections show that we can use the direction of the induced electric field vector as a tool to study the mechanism for the radio emission from air showers. In addition to the geomagnetic emission process which leads to an electric field vector pointing in a direction which is fixed by the incoming direction of the cosmic ray and the magnetic field vector of the Earth, there is another electric field component which is pointing radially towards the core of the shower. For the present equipment sited at the Pierre Auger Observatory and for the set of showers observed, this radial component has on average a relative strength of $(14 \pm 2)$\% with respect to the component induced by the geomagnetic emission process and it is pointing towards the core of the shower. These results are supported by the analysis of the data obtained by the prototype, which is presented in Appendix \ref{app:MAXIMA}.

\clearpage
\section{Comparison with calculations}\label{sec:comparison}

In this section, we compare the results shown in Section \ref{sec:r_analysis} with simulations using different approaches listed in Table \ref{tab:models}. The codes CoREAS \cite{Refworks:223}, EVA1.01 \cite{Refworks:212}, REAS3.1 \cite{Refworks:11, Refworks:224}, SELFAS \cite{Refworks:213}, and ZHAireS \cite{Refworks:220} use a multiple-layered structure for the atmosphere, whereas MGMR \cite{Refworks:3} has a single layer. All of these models assume an exponential profile for the density of the air (denoted as $\rho$ in this table) per layer. The treatment of the index of refraction differs from model to model: SELFAS and MGMR have a constant index of refraction (equal to unity); for the other models, the index of refraction follows the density of air. Here we note that recently SELFAS has been updated to include an index of refraction wihich differs from unity. Another difference between the codes is the description of the shower development; they use either a parameterized model for the development of the particle density distribution within the shower (EVA1.01, MGMR, and SELFAS) or they make a realistic Monte Carlo calculation to obtain these density distributions (CoREAS, REAS3.1, ZHAireS).

\begin{table}[h]
\caption{\label{tab:models}
    Characterization of the simulation codes.}
\begin{ruledtabular}
\begin{tabular}{lllll}
    model      &Ref.                       &layers         &refractivity         &interaction model      \\ \hline
    CoREAS    &\cite{Refworks:223}        &multiple       &$\propto \rho$           &Monte-Carlo               \\
    EVA1.01   &\cite{Refworks:212}        &multiple       &$\propto \rho$ (see \cite{Refworks:212}) &parameterized\\
    MGMR      &\cite{Refworks:3}          &single         &0                        &parameterized                          \\
    REAS3.1   &\cite{Refworks:11}         &multiple       &$\propto \rho$           &Monte-Carlo               \\
    SELFAS    &\cite{Refworks:213}        &multiple       &0                        &parameterized               \\
    ZHAireS   &\cite{Refworks:220}        &multiple       &$\propto \rho$           &Monte-Carlo               \\
\end{tabular}
\end{ruledtabular}
\end{table}

The comparison between the measured data and the values predicted by the various models was done with the analysis package \cite{Refworks:32,Refworks:179}, which was introduced in Section \ref{sec:rdanalysis}. The data from the SD together with the position and orientation of each RDS were used to predict the electric field strength in each polarization direction of an RDS. The full response of the RDS (the response of the analog chain and the antenna gain) was then used to predict the value of $R$ (see Eq. (\ref{eq:observable})), here denoted as $R\mbox{(pr.)}$. We started from the predicted electric field strength at the position of an RDS. As a next step in this chain, these values led to predicted values at the voltage level, very similar to the ones obtained in the actual measurement. Once these values were obtained, the same scheme was followed as the one used for the analysis of the experimental data, leading to predicted values for $R$. To estimate the uncertainty in this prediction, we performed Monte Carlo simulations for 25 different showers, all with slightly different shower parameters, where we used the estimated uncertainties from the SD analysis for each of the shower parameters and their correlations. The ensemble of these 25 predicted values for $R$ was used to determine the uncertainty denoted as $\sigma\mbox{(pr.)}$.

\begin{figure}[h!t]
	\centering
        \subfigure{\includegraphics[width=0.45\textwidth, viewport=0 0 450 450, clip]{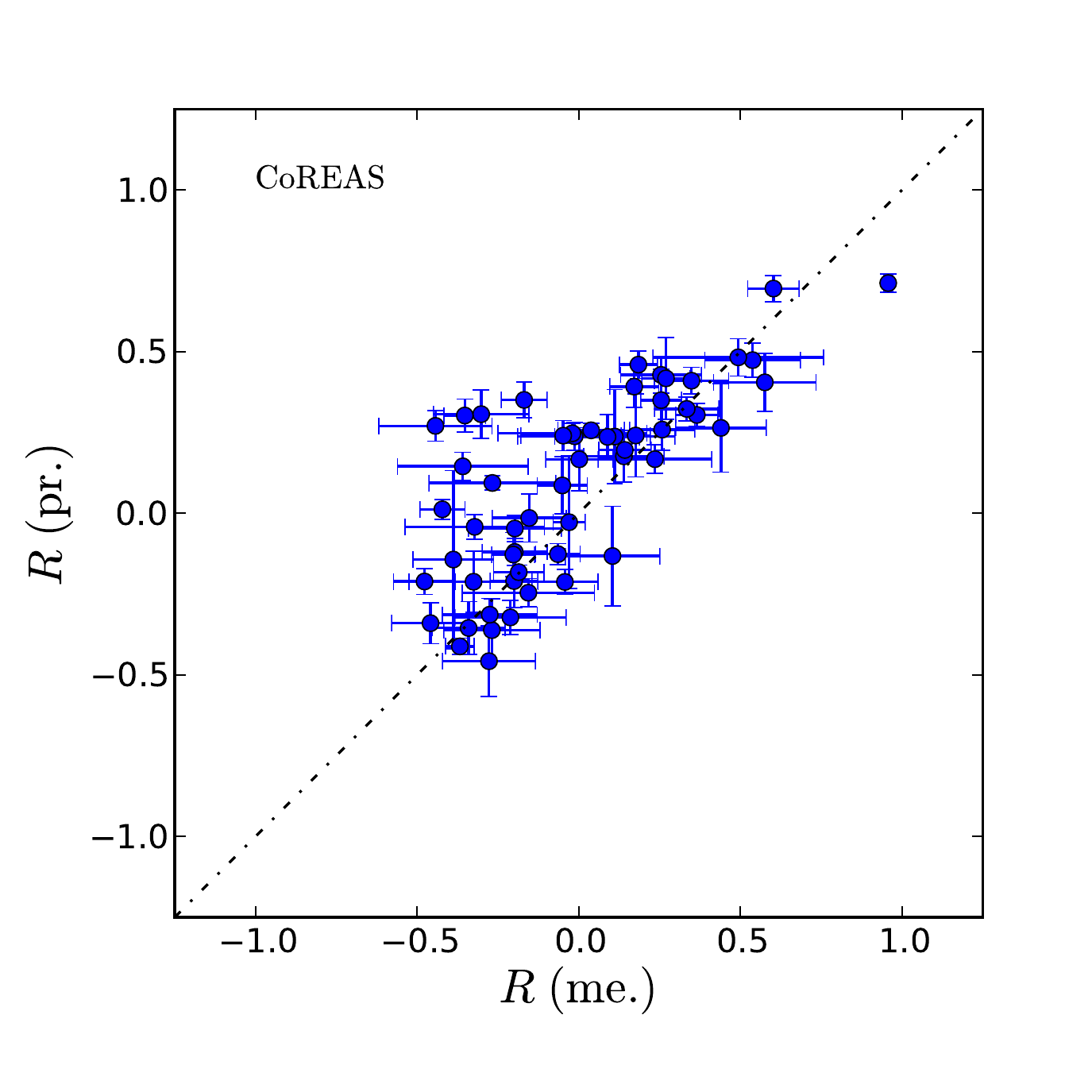}}
        \hspace{.1cm}
		\subfigure{\includegraphics[width=0.45\textwidth, viewport=0 0 450 450, clip]{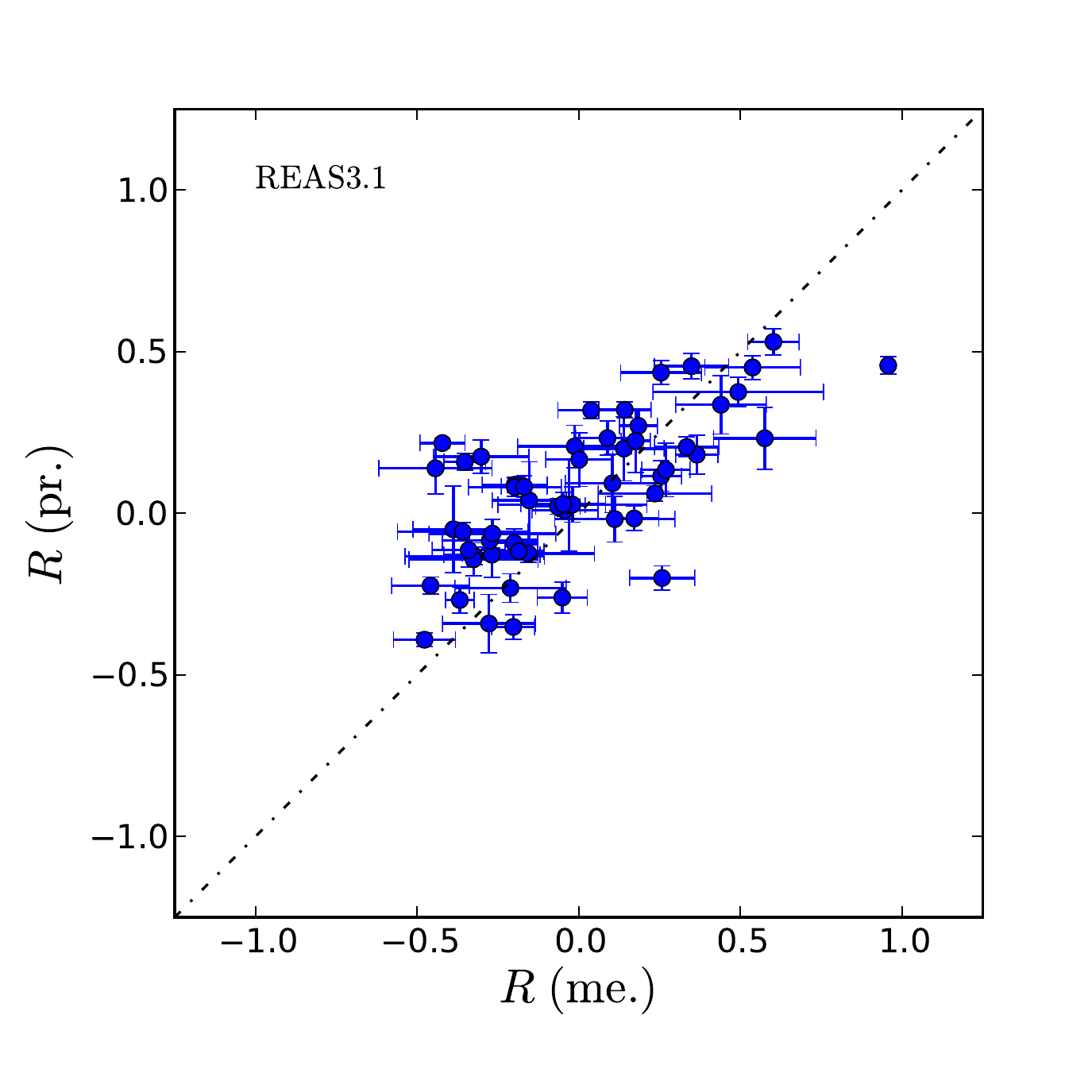}}
        \\ \vspace{-1cm}
		\subfigure{\includegraphics[width=0.45\textwidth, viewport=0 0 450 450, clip]{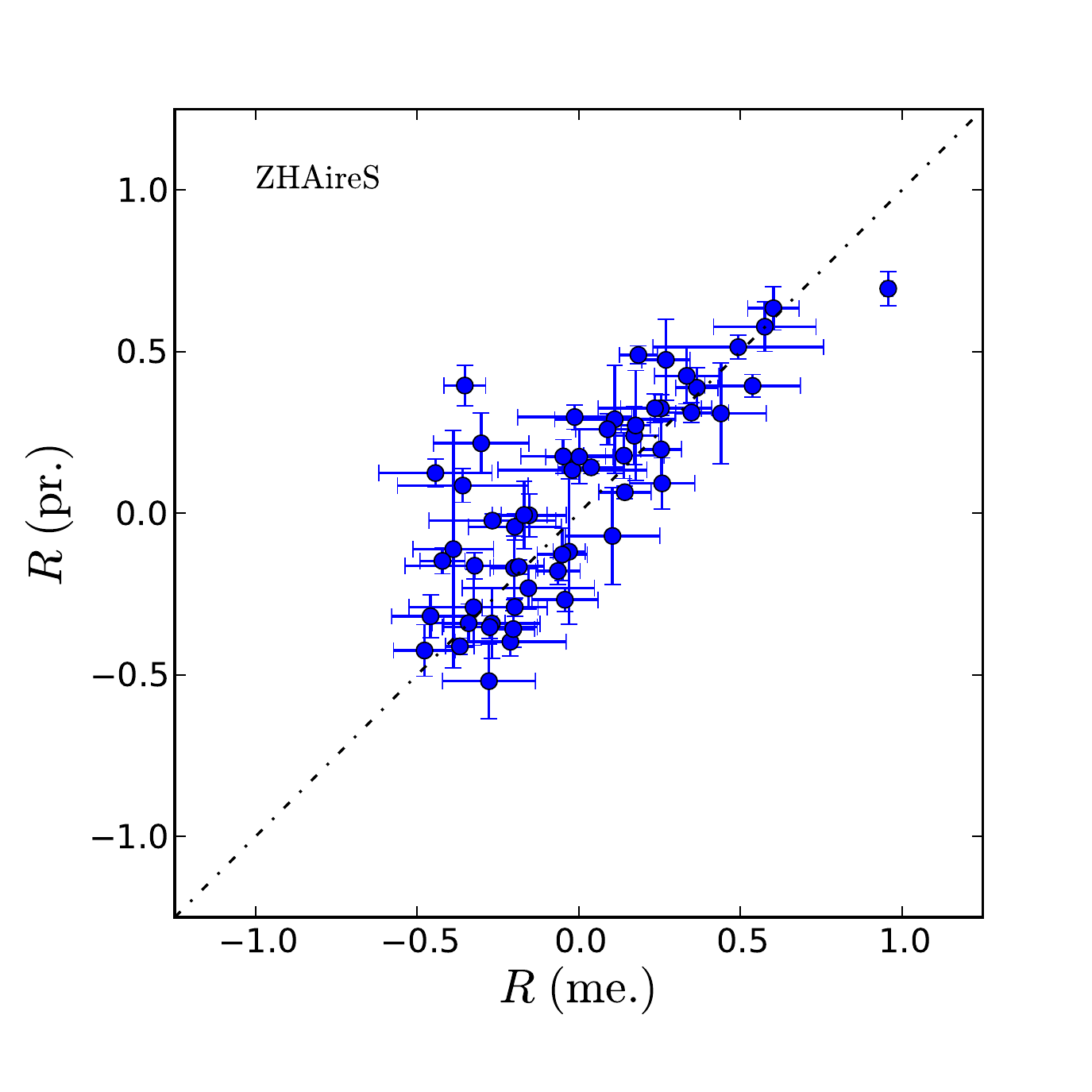}}
        \hspace{.1cm}
		\subfigure{\includegraphics[width=0.45\textwidth, viewport=0 0 450 450, clip]{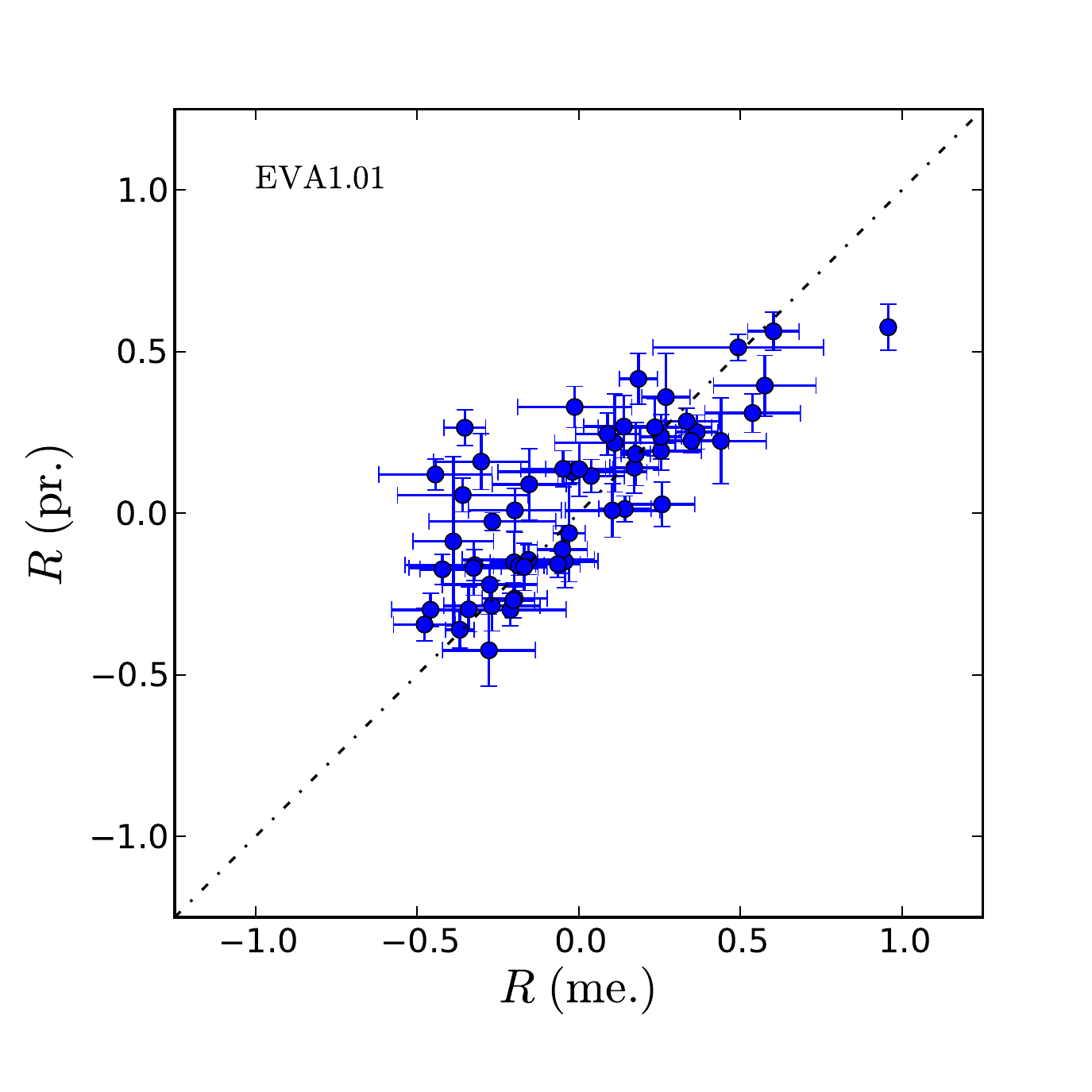}}
        \\ \vspace{-1cm}
		\subfigure{\includegraphics[width=0.45\textwidth, viewport=0 0 450 450, clip]{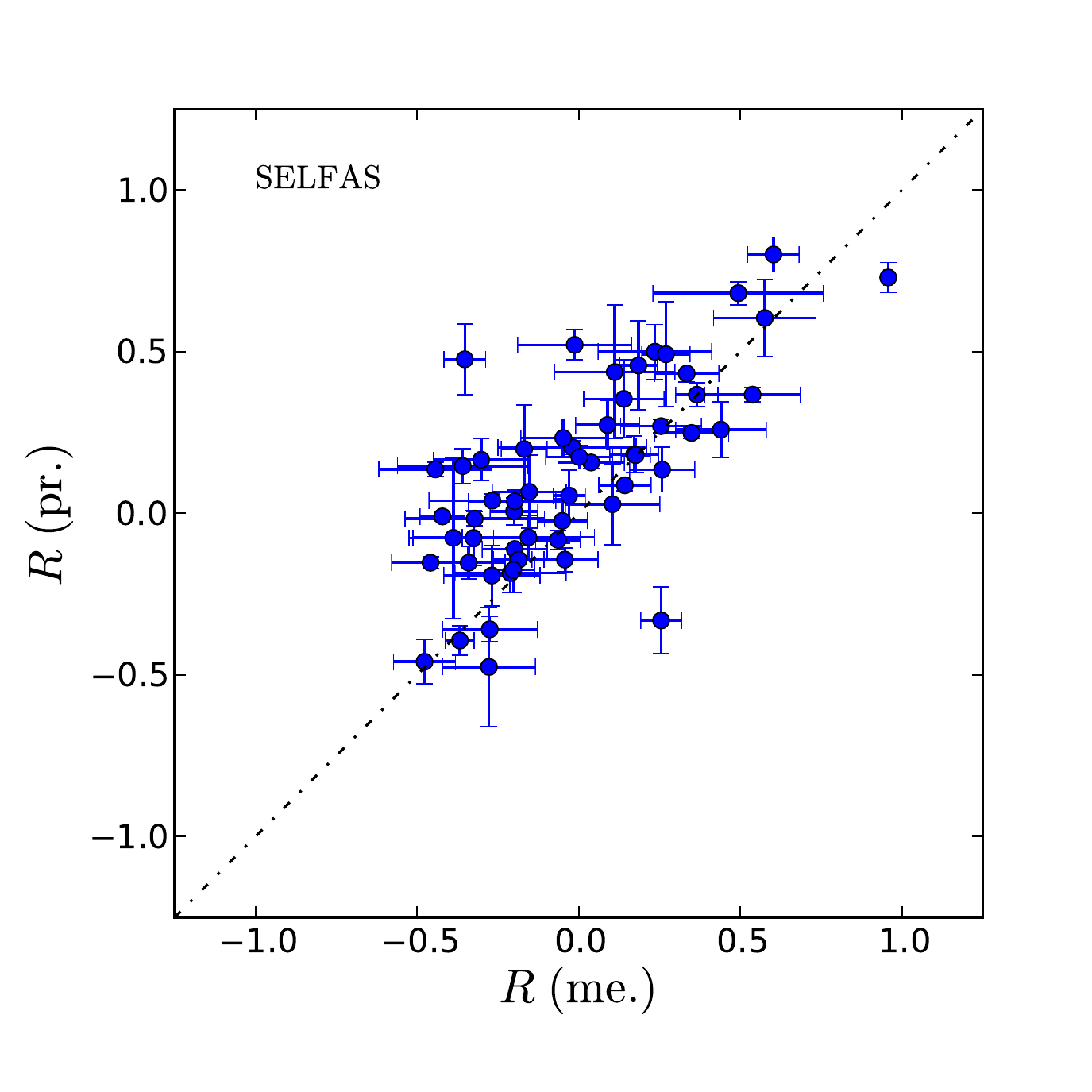}}
        \hspace{.1cm}
		\subfigure{\includegraphics[width=0.45\textwidth, viewport=0 0 450 450, clip]{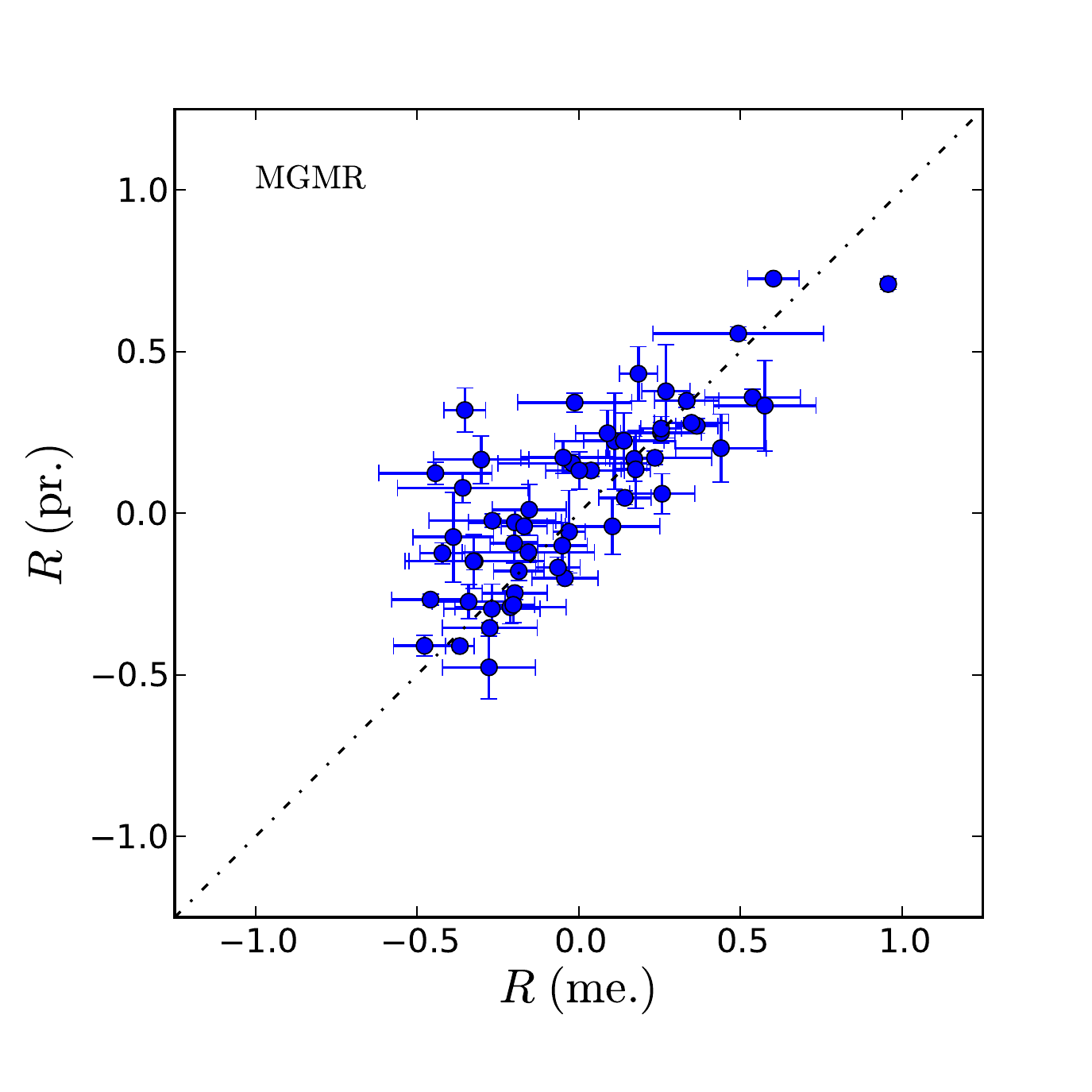}}
    \caption{Predicted versus measured values of the parameter $R$; see text for details.}
	\label{fig:RversusRAERA}
\end{figure}

\begin{figure}[h!t]
	\centering
		 \subfigure{\includegraphics[width=0.45\textwidth, viewport=0 0 450 450, clip]{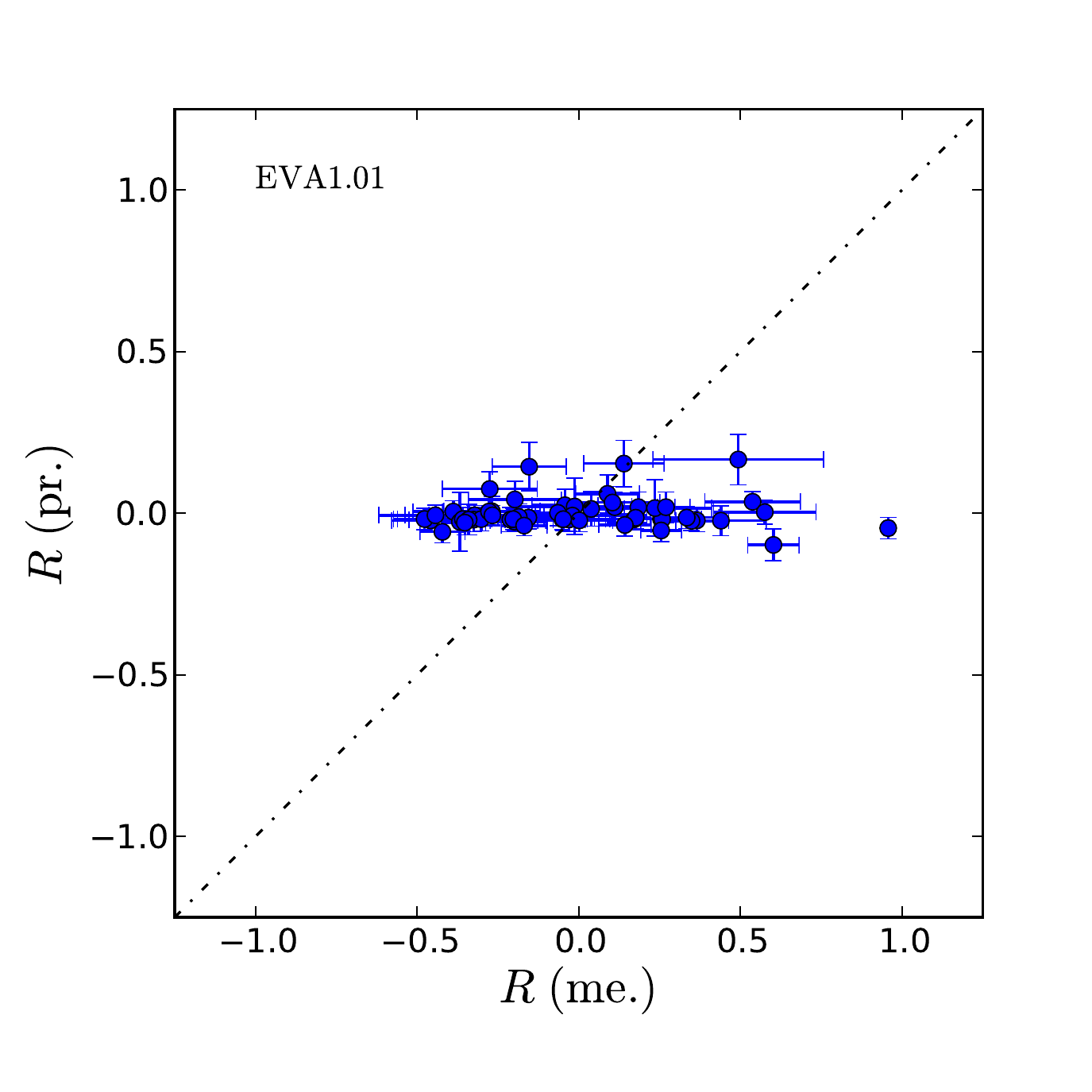}}
        \hspace{.1cm}
		 \subfigure{\includegraphics[width=0.45\textwidth, viewport=0 0 450 450, clip]{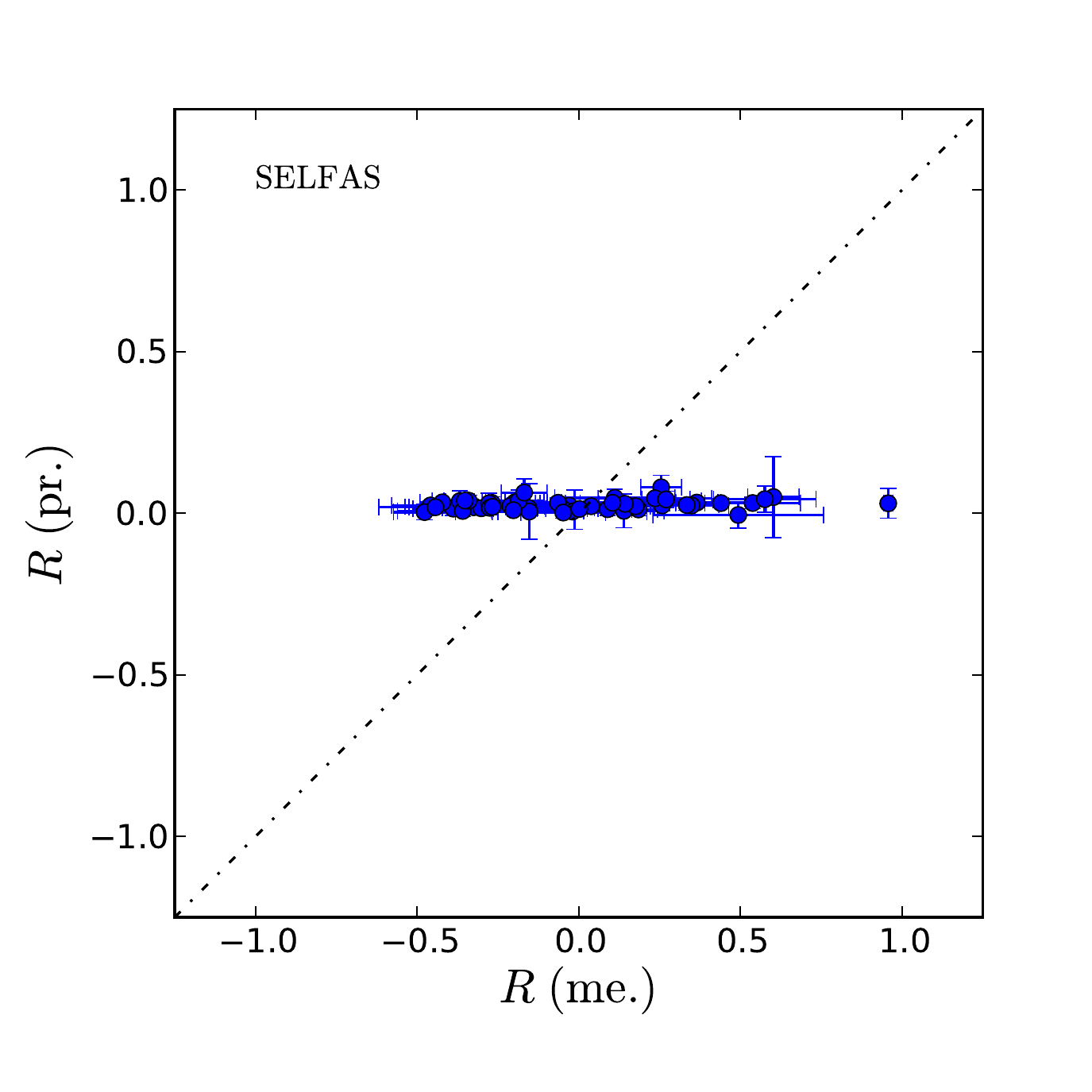}}
    \caption{Measured versus predicted values of the parameter $R$ in two cases where the charge-excess component has been switched off in the simulations.}
	\label{fig:Rversus0}
\end{figure}

Figure \ref{fig:RversusRAERA} shows for the AERA24 data set the comparison between the measured values of $R$ versus those predicted. For each of the six models, the Pearson correlation coefficient $\rho_P$ between the data and predictions and its associated asymmetric 95\% confidence range indicated by the lower $\rho_{L}$ (higher $\rho_{H}$) limit are listed in Table \ref{tab:pearsonAERA}. These coefficients are typically $0.7$. For some of these approaches, it is possible to explicitly switch off the contributions caused by the charge-excess process, which leads to values of $R\mbox{(pr.)}$ which are close to zero. Examples of such calculations are shown in Fig. \ref{fig:Rversus0}. Also in this case the correlation coefficients have been calculated and are listed in Table \ref{tab:pearsonAERA}. In this case the correlation coefficients are close to zero.

\begin{table}[h]
\caption{\label{tab:pearsonAERA} Pearson correlation coefficients $\rho_P$ and their 95\% confidence ranges.}
\begin{ruledtabular}
\begin{tabular}{l  lll lll}
               &\multicolumn{3}{c}{charge excess}      &\multicolumn{3}{c}{no charge excess} \\ \hline
    model       &$\rho_{L}$ &$\rho_P$     &$\rho_{H}$    &$\rho_{L}$  &$\rho_P$     &$\rho_{H}$         \\ \hline
    CoREAS      &0.58   &0.67   &0.75   &      &      &        \\
    EVA1.01     &0.60   &0.70   &0.78   &-0.16  &0.04   &0.24     \\
    MGMR        &0.62   &0.71   &0.78   &-0.20  &-0.01   &0.20     \\
    REAS3.1     &0.54   &0.63   &0.71   &       &      &        \\
    SELFAS      &0.55   &0.64   &0.72   &-0.22  &0.09   &0.37     \\
    ZHAireS     &0.61   &0.70   &0.78   &       &      &        \\
\end{tabular}
\end{ruledtabular}
\end{table}

It is seen from Table \ref{tab:pearsonAERA}, that the inclusion of the charge-excess emission improves significantly the value of the correlation $\rho_P$. We also calculated the reduced $\chi^2$ values, which are defined as:
\begin{equation}
    \chi^{2}=\sum_{i=1}^{D}\frac{\left[ R_{i}\mbox{(me.)} - R_{i}\mbox{(pr.)}    \right]^{2} }
     {\sigma_{i}^2\mbox{(me.)} + \sigma_{i}^2 \mbox{(pr.)}}\label{eq:chi2idef}.
\end{equation}

The calculated reduced $\chi^2$ values are about equal to 3 if the charge-excess effect is included and roughly equal to 20 in case the contributions of charge-excess emission have been switched off. Therefore, although the inclusion of the charge-excess contribution clearly improves the correlation coefficients as well as the reduced $\chi^2$ values, the present data set can not be fully described by these calculations. Furthermore, the various models produce slightly different results, which, in itself, is very interesting and could lead to further insights into the modeling of emission processes by air showers. However, such a discussion goes beyond the scope of the present paper. For completeness we present in Appendix \ref{app:MAXIMA} the results of the comparison between model calculations and the data obtained with the prototype.

\clearpage
\section{Conclusion}\label{sec:conclusion}

We have studied with two different radio-detection setups deployed at the Pierre Auger Observatory the emission around 50 MHz of radio waves from air showers. For a sample of 37 air showers the electric field strength has been analyzed as a tool to disentangle the emission mechanism caused by the geomagnetic and the charge-excess processes. For the present data sets, the emission is dominated by the geomagnetic emission process while, in addition, a significant fraction of on average $(14 \pm 2)\%$ is attributed to a radial component which is consistent with the charge-excess emission mechanism. Detailed simulations have been performed where both emission processes were included. The comparison of these simulations with the data underlines the importance of including the charge-excess mechanism in the description of the measured data. However, further refinements of the models might be required to fully describe the present data set. A possible reason for the incomplete description of the data by the models might be an underestimate of (systematic) errors in the data sets or the effect of strong electric fields in the atmosphere.

The successful installation, commissioning, and operation of the Pierre Auger Observatory
would not have been possible without the strong commitment and effort
from the technical and administrative staff in Malarg\"ue.

We are very grateful to the following agencies and organizations for financial support:
Comisi\'on Nacional de Energ\'ia At\'omica,
Fundaci\'on Antorchas,
Gobierno De La Provincia de Mendoza,
Municipalidad de Malarg\"ue,
NDM Holdings and Valle Las Le\~nas, in gratitude for their continuing
cooperation over land access, Argentina;
the Australian Research Council;
Conselho Nacional de Desenvolvimento Cient\'ifico e Tecnol\'ogico (CNPq),
Financiadora de Estudos e Projetos (FINEP),
Funda\c{c}\~ao de Amparo \`a Pesquisa do Estado de Rio de Janeiro (FAPERJ),
S\~ao Paulo Research Foundation (FAPESP) Grants \#2010/07359-6, \#1999/05404-3,
Minist\'erio de Ci\^{e}ncia e Tecnologia (MCT), Brazil;
AVCR, MSMT-CR LG13007, 7AMB12AR013, MSM0021620859, and TACR TA01010517 , Czech Republic;
Centre de Calcul IN2P3/CNRS,
Centre National de la Recherche Scientifique (CNRS),
Conseil R\'egional Ile-de-France,
D\'epartement  Physique Nucl\'eaire et Corpusculaire (PNC-IN2P3/CNRS),
D\'epartement Sciences de l'Univers (SDU-INSU/CNRS), France;
Bundesministerium f\"ur Bildung und Forschung (BMBF),
Deutsche Forschungsgemeinschaft (DFG),
Finanzministerium Baden-W\"urttemberg,
Helmholtz-Gemeinschaft Deutscher Forschungszentren (HGF),
Ministerium f\"ur Wissenschaft und Forschung, Nordrhein-Westfalen,
Ministerium f\"ur Wissenschaft, Forschung und Kunst, Baden-W\"urttemberg, Germany;
Istituto Nazionale di Fisica Nucleare (INFN),
Ministero dell'Istruzione, dell'Universit\`a e della Ricerca (MIUR),
Gran Sasso Center for Astroparticle Physics (CFA), CETEMPS Center of Excellence, Italy;
Consejo Nacional de Ciencia y Tecnolog\'ia (CONACYT), Mexico;
Ministerie van Onderwijs, Cultuur en Wetenschap,
Nederlandse Organisatie voor Wetenschappelijk Onderzoek (NWO),
Stichting voor Fundamenteel Onderzoek der Materie (FOM), Netherlands;
Ministry of Science and Higher Education,
Grant Nos. N N202 200239 and N N202 207238,
The National Centre for Research and Development Grant No ERA-NET-ASPERA/02/11, Poland;
Portuguese national funds and FEDER funds within COMPETE - Programa Operacional Factores de Competitividade through
Funda\c{c}\~ao para a Ci\^{e}ncia e a Tecnologia, Portugal;
Romanian Authority for Scientific Research ANCS,
CNDI-UEFISCDI partnership projects nr.20/2012 and nr.194/2012,
project nr.1/ASPERA2/2012 ERA-NET, PN-II-RU-PD-2011-3-0145-17, and PN-II-RU-PD-2011-3-0062, Romania;
Ministry for Higher Education, Science, and Technology,
Slovenian Research Agency, Slovenia;
Comunidad de Madrid,
FEDER funds,
Ministerio de Ciencia e Innovaci\'on and Consolider-Ingenio 2010 (CPAN),
Xunta de Galicia, Spain;
The Leverhulme Foundation,
Science and Technology Facilities Council, United Kingdom;
Department of Energy, Contract Nos. DE-AC02-07CH11359, DE-FR02-04ER41300, DE-FG02-99ER41107,
National Science Foundation, Grant No. 0450696,
The Grainger Foundation USA;
NAFOSTED, Vietnam;
Marie Curie-IRSES/EPLANET, European Particle Physics Latin American Network,
European Union 7th Framework Program, Grant No. PIRSES-2009-GA-246806;
and UNESCO.

\clearpage
\appendix
\section{Bias on the determination of $R$}\label{app:A}

We introduced in section \ref{sec:r_analysis} Eq. (\ref{eq:observable}) to determine the value of $R$. In case the signals are of low amplitude this calculation will be affected by the magnitude of noise. In order to remove this systematic effect, one needs to subtract this noise. The equation for the determination of the observable $R$ in the presence of noise involves yet another transformation, leading to a slightly more complex formula as compared to Eq. (\ref{eq:observable}):

\begin{equation}
R(\psi)=\frac{\sum_{j=1}^{25}(|\mathcal{E}_{j+k,\lambda}|^{2}-|\mathcal{E}_{j+k,\rho}|^{2})/25-\sum_{j=1}^{320}(|\mathcal{E}_{j+m_{1},\lambda}|^{2}-|\mathcal{E}_{j+m_{1},\rho}|^{2})/320}{\sum_{j=1}^{25}(|\mathcal{E}_{j+k,\lambda}|^{2}+|\mathcal{E}_{j+k,\rho}|^{2})/25-\sum_{j=1}^{320}(|\mathcal{E}_{j+m_{1},\lambda}|^{2}+|\mathcal{E}_{j+m_{1},\rho}|^{2})/320}
\end{equation}

The indices $\lambda$ and $\rho$ are due to a coordinate transformation on the data such that $\mathcal{E}_{\lambda} = (\mathcal{E}_{\xi} + \mathcal{E}_{\eta} ) / \sqrt{2}$ and $\mathcal{E}_{\rho} = (\mathcal{E}_{\xi} - \mathcal{E}_{\eta} ) / \sqrt{2}$. For the present data sets, the correction to the denominator was most significant, whereas the correction to the numerator was smaller and took care of possible differences in the noise levels in the coordinate system defined by the variables $\lambda$ and $\rho$.

\section{Extraction of the error on $a$}\label{app:tech_a}

In Section \ref{sec:a_analysis} we have presented the definition of the parameter $a$. Here we explain how to estimate the error on the value of $a$ for each individual data point. First we estimate the $p$-value of measuring $\phi_p$ for a given value of $a$, $p(\phi_a|a)$, where $a$ ranges from $-1.0$ to $+1.0$. This probability was obtained by generating a probability density function $f(\phi'_a|a)$ of polarization angles using Eq. (\ref{eq:pre_pol_ang2}). This probability density function $f$ was obtained by varying the location and direction of the shower axis according to their uncertainties, varying the orientation of the geomagnetic field according to its uncertainty, and adding a random angle that is distributed according to the measurement uncertainty on the polarization angle. From the function $f$ we calculated the most probable value for $a$ and the 68\% uncertainty, as this is indicated with the black error bars in Fig. \ref{fig:a-aera}.

\section{Polarization data from the prototype}\label{app:MAXIMA}

\begin{figure}[t]
    \centering
    \includegraphics[height=0.7\textwidth]{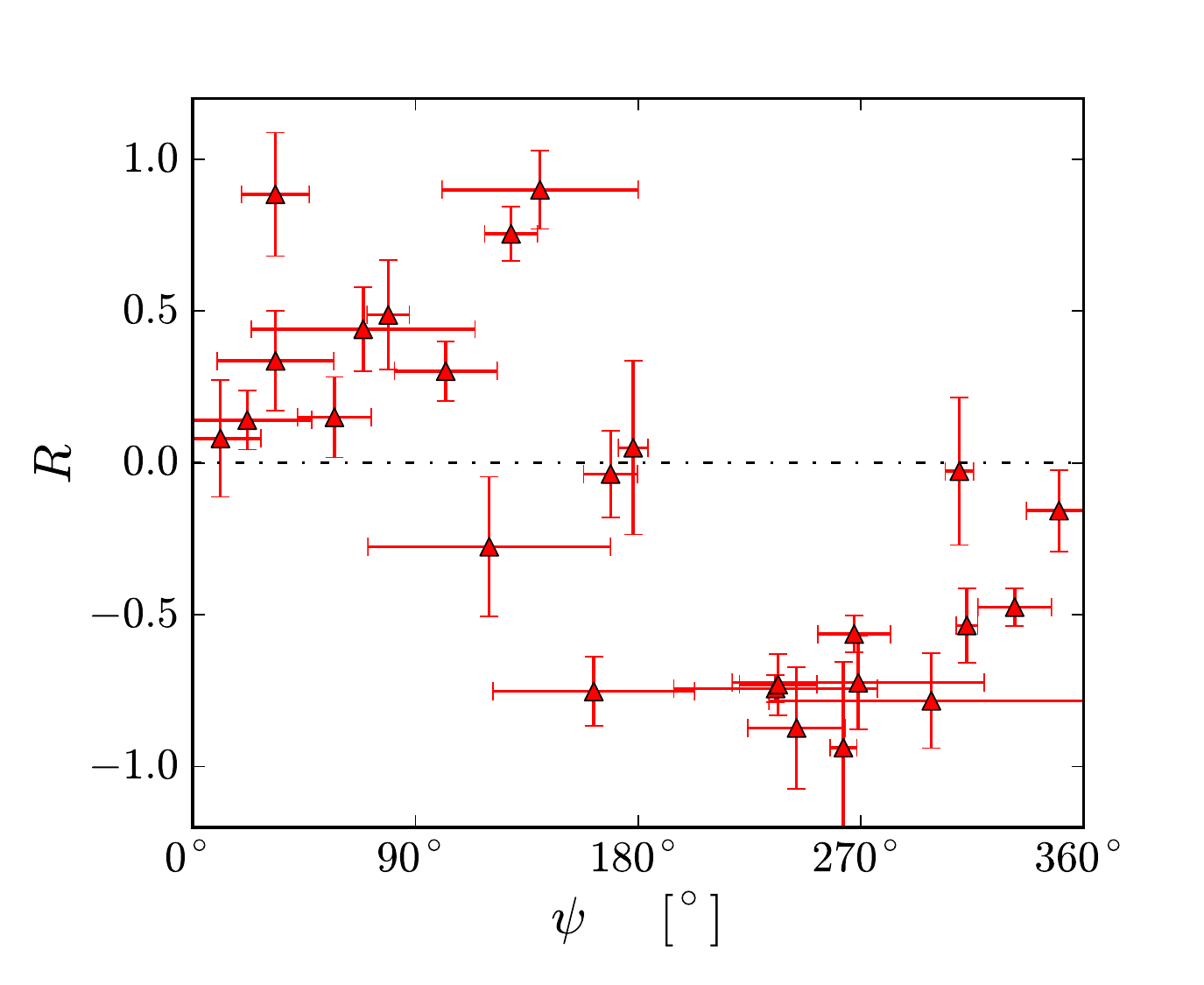}
    \caption{The calculated value of $R$ and its uncertainty as a function of the observation angle $\psi$. See also the caption to Fig. \ref{fig:Observer-angle-dependenceAERA}.}
    \label{fig:Observer-angle-dependenceMAXI}
\end{figure}

\begin{figure}[ht!]
    \centering
    \subfigure{\includegraphics[width=.49\textwidth]{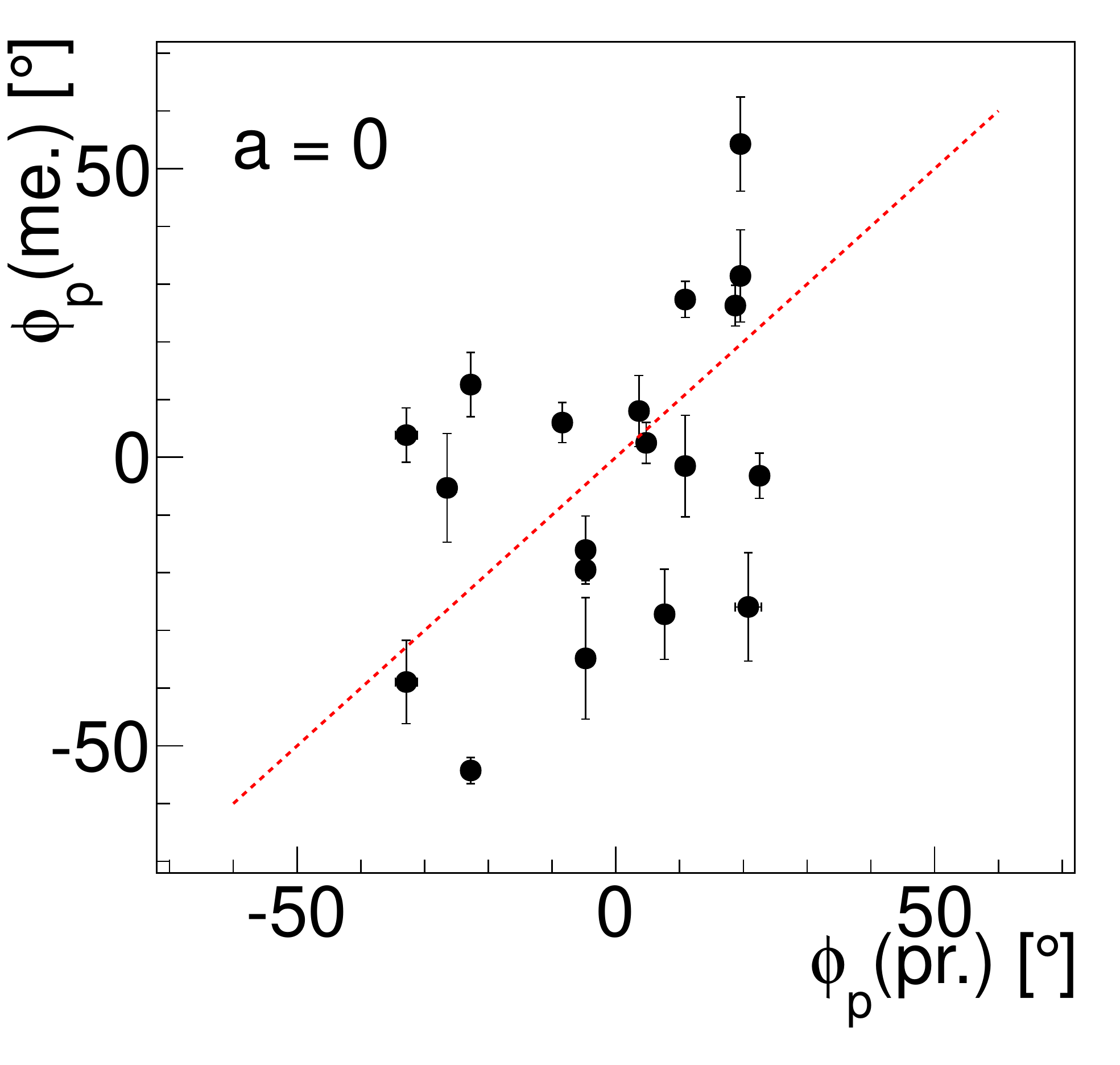}}
    \subfigure{\includegraphics[width=.49\textwidth]{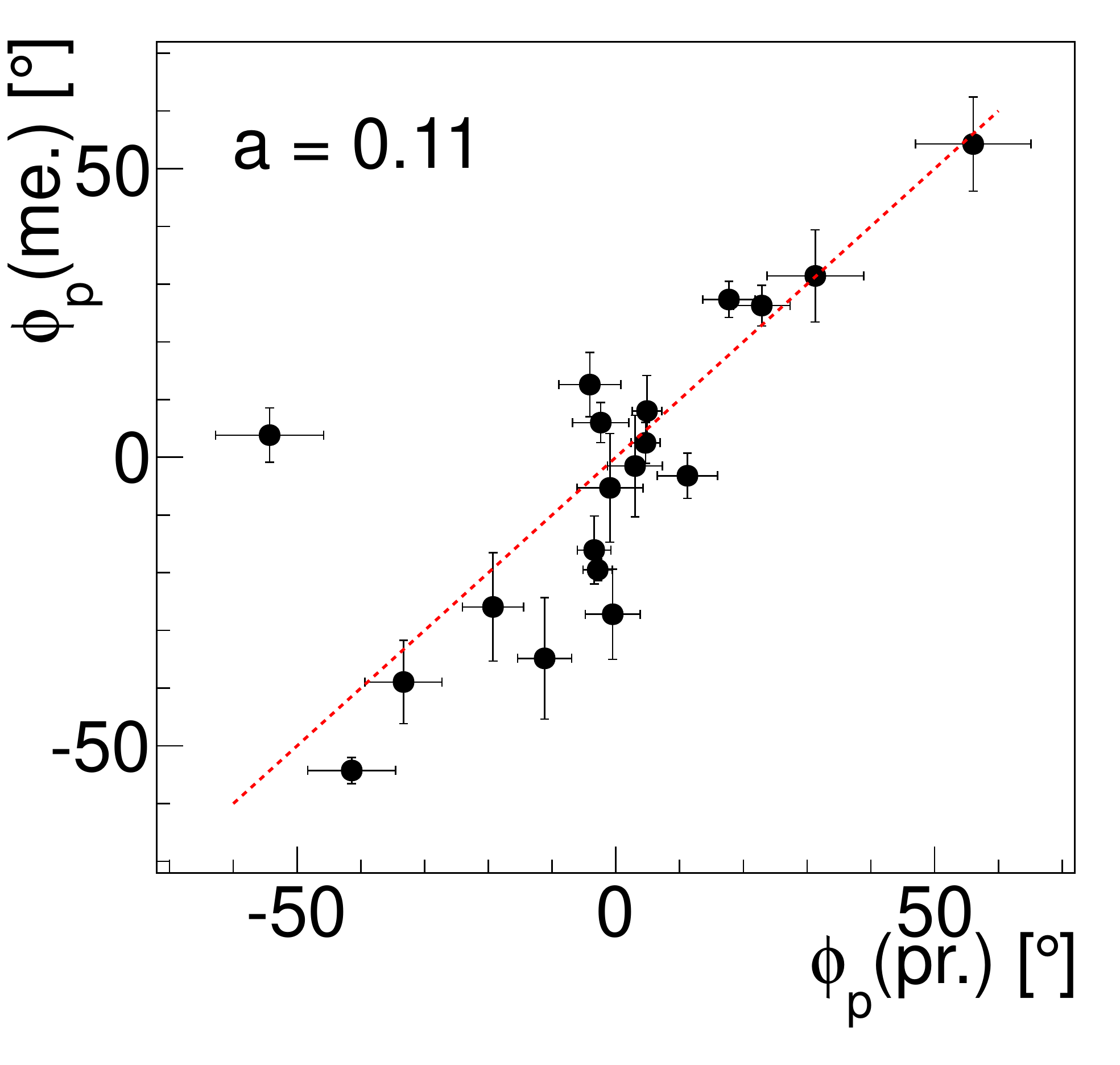}}
    \caption{Left panel: the measured polarization angle for pure geomagnetic emission, $a = 0$ (see Eqs. (\ref{eq:pre_pol_ang2}) and (\ref{eq:definitionofa})), versus the predicted polarization angle for the data set recorded with the prototype. Right panel: the same data set but for a value of $a = 0.11 \pm 0.07$. See also the captions to Figs. \ref{fig:polarizationAERA0} and \ref{fig:polAng-a0123}.}
   \label{fig:polarizationMAXI}
\end{figure}

\begin{figure}[h!t]
	\centering
        \subfigure{\includegraphics[width=0.45\textwidth, viewport=0 0 450 450, clip]{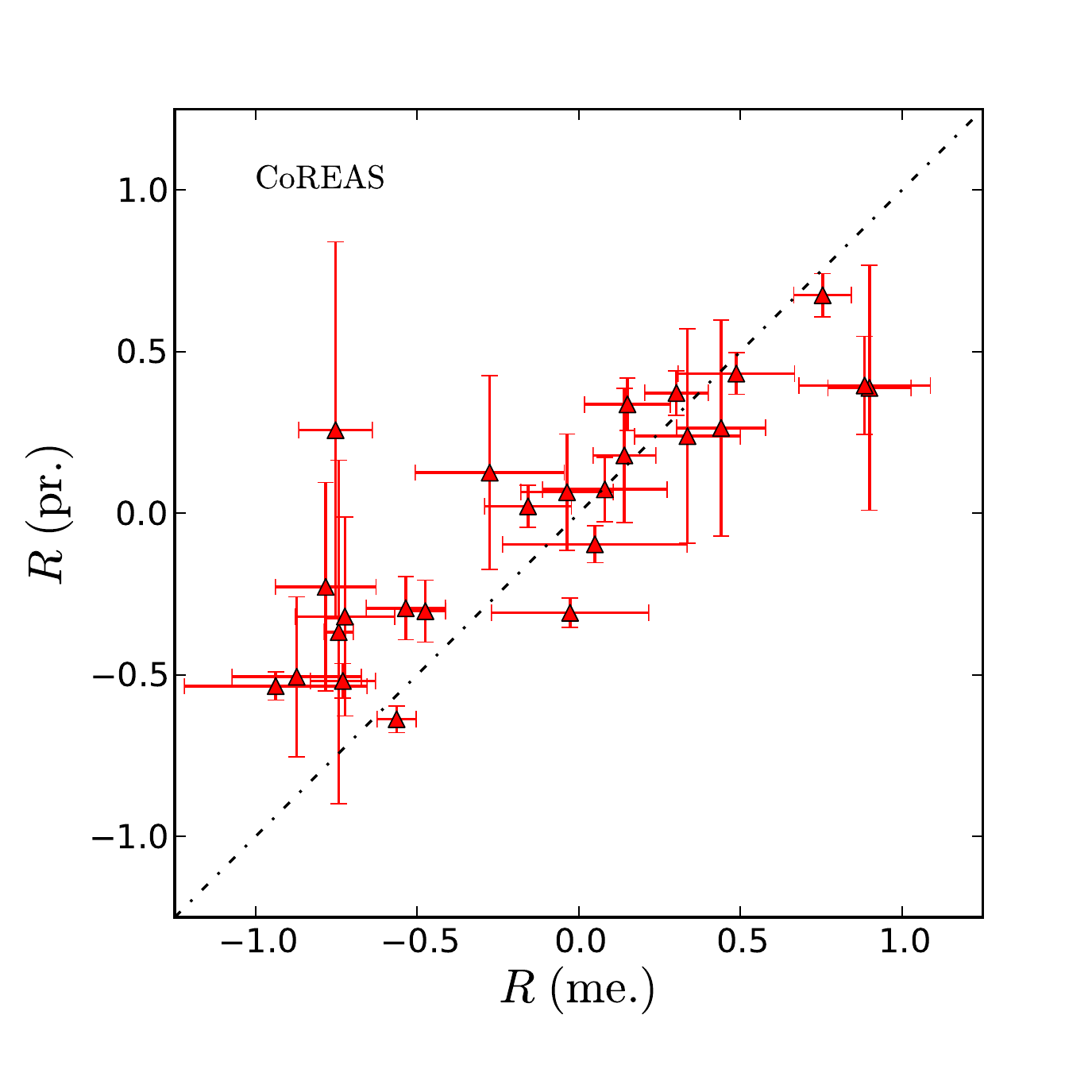}}
        \hspace{.1cm}
		\subfigure{\includegraphics[width=0.45\textwidth, viewport=0 0 450 450, clip]{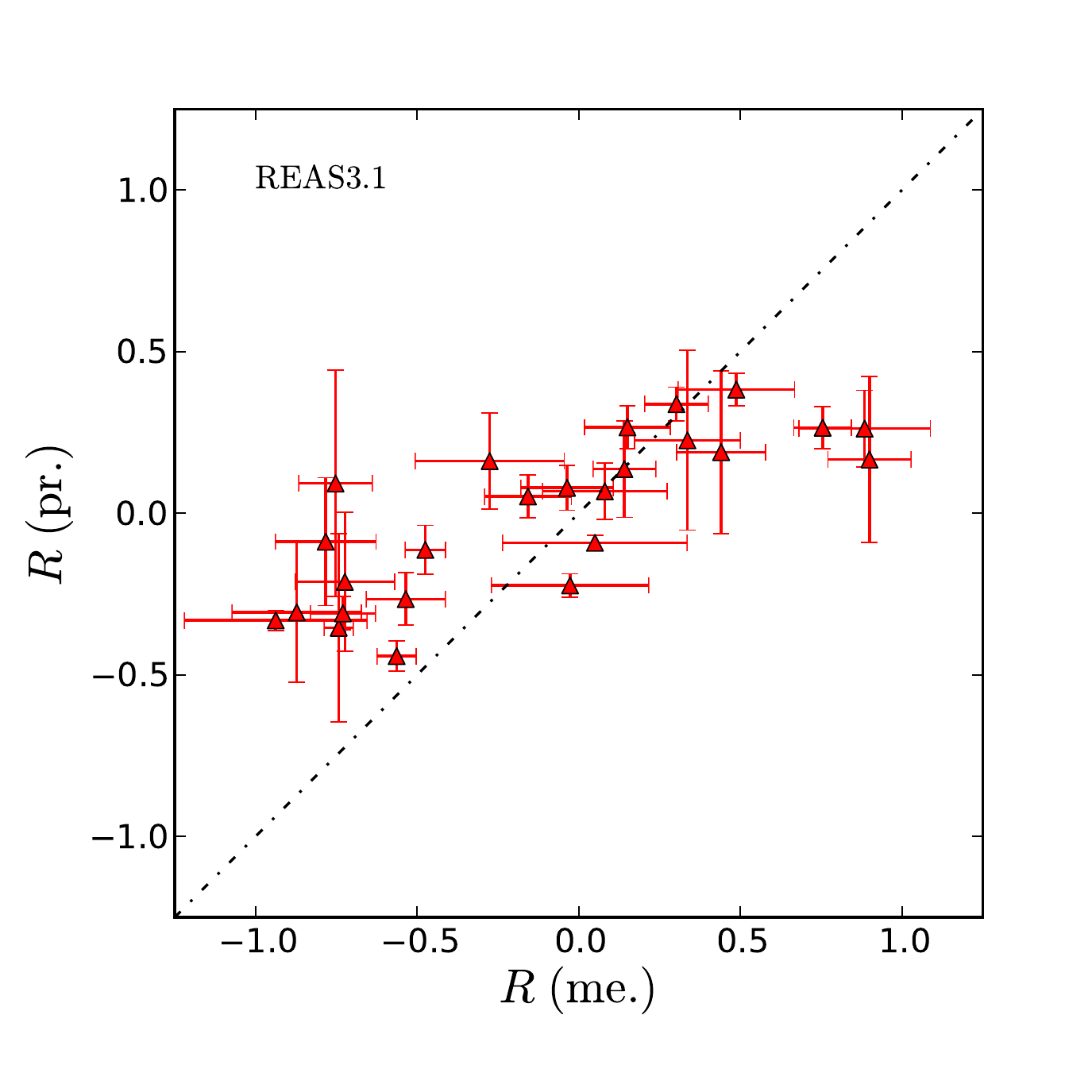}}
        \\ \vspace{-1cm}
		\subfigure{\includegraphics[width=0.45\textwidth, viewport=0 0 450 450, clip]{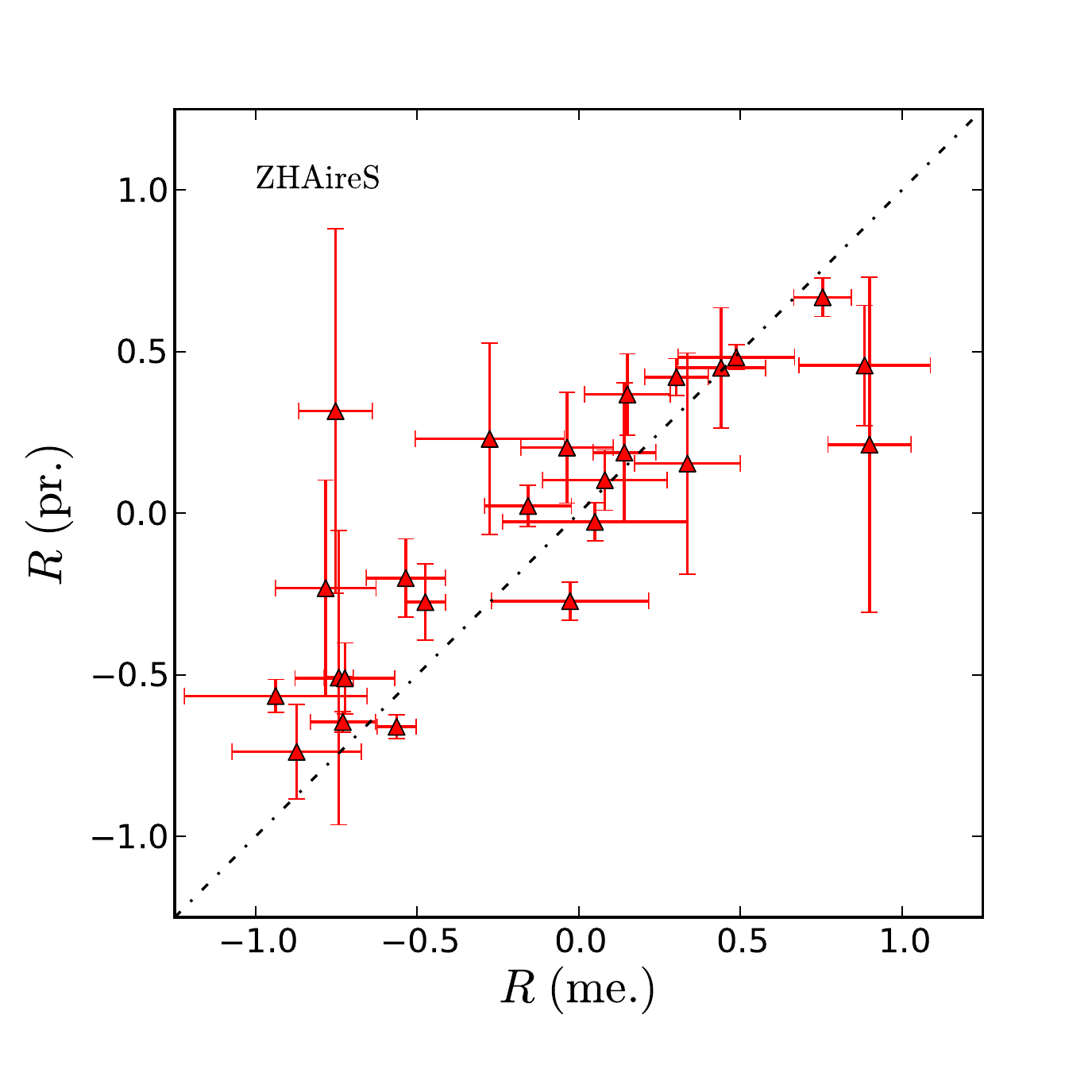}}
        \hspace{.1cm}
		\subfigure{\includegraphics[width=0.45\textwidth, viewport=0 0 450 450, clip]{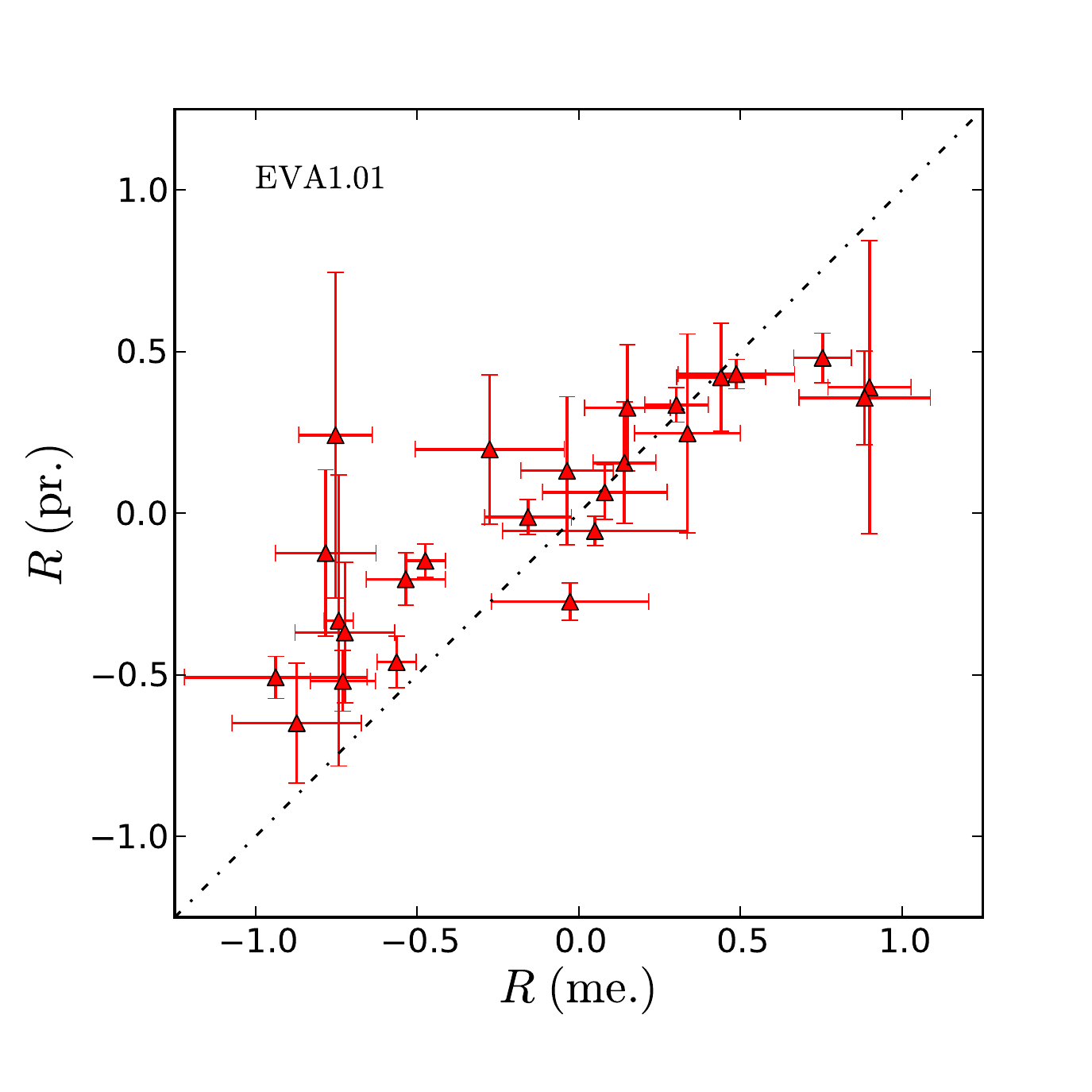}}
        \\ \vspace{-1cm}
		\subfigure{\includegraphics[width=0.45\textwidth, viewport=0 0 450 450, clip]{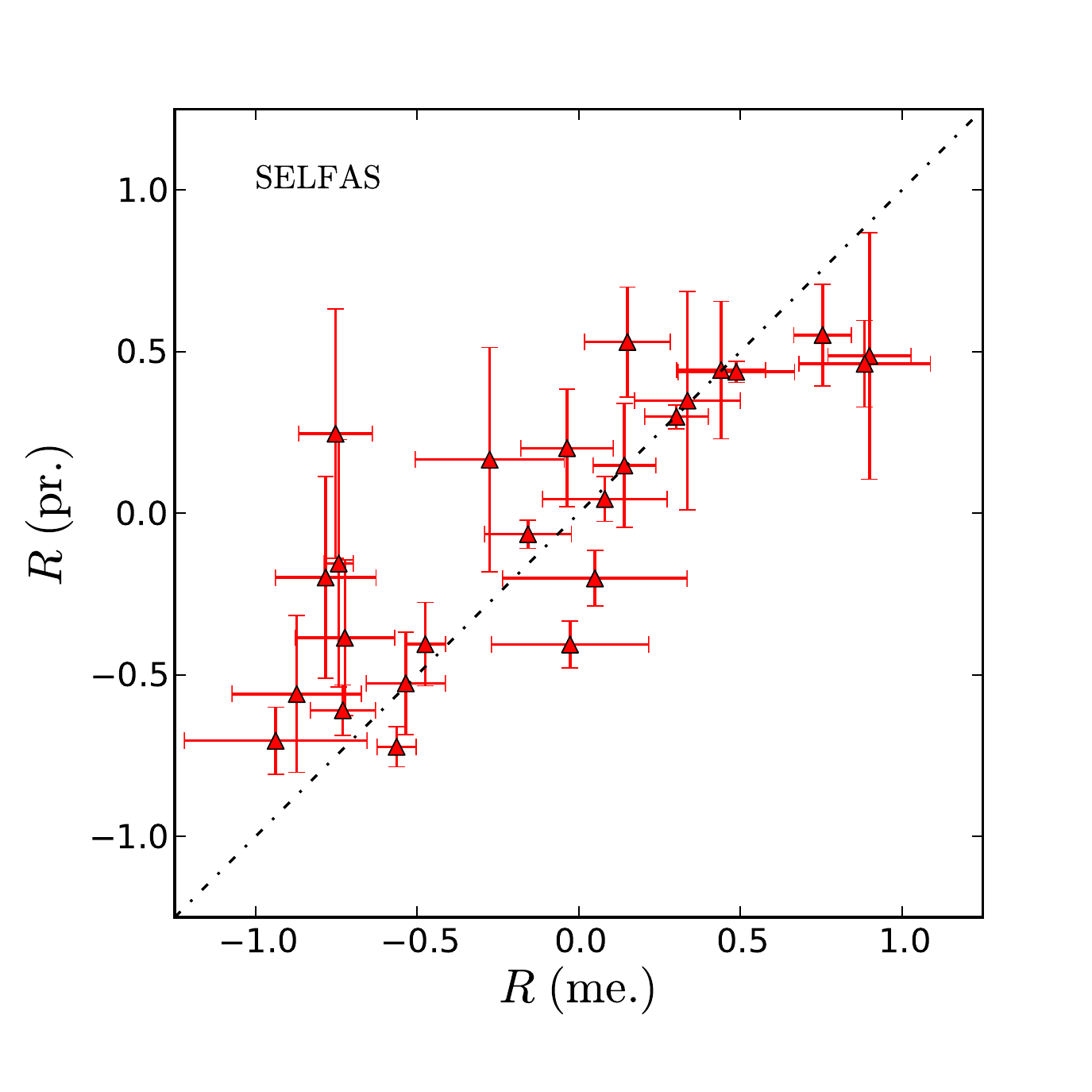}}
        \hspace{.1cm}
		\subfigure{\includegraphics[width=0.45\textwidth, viewport=0 0 450 450, clip]{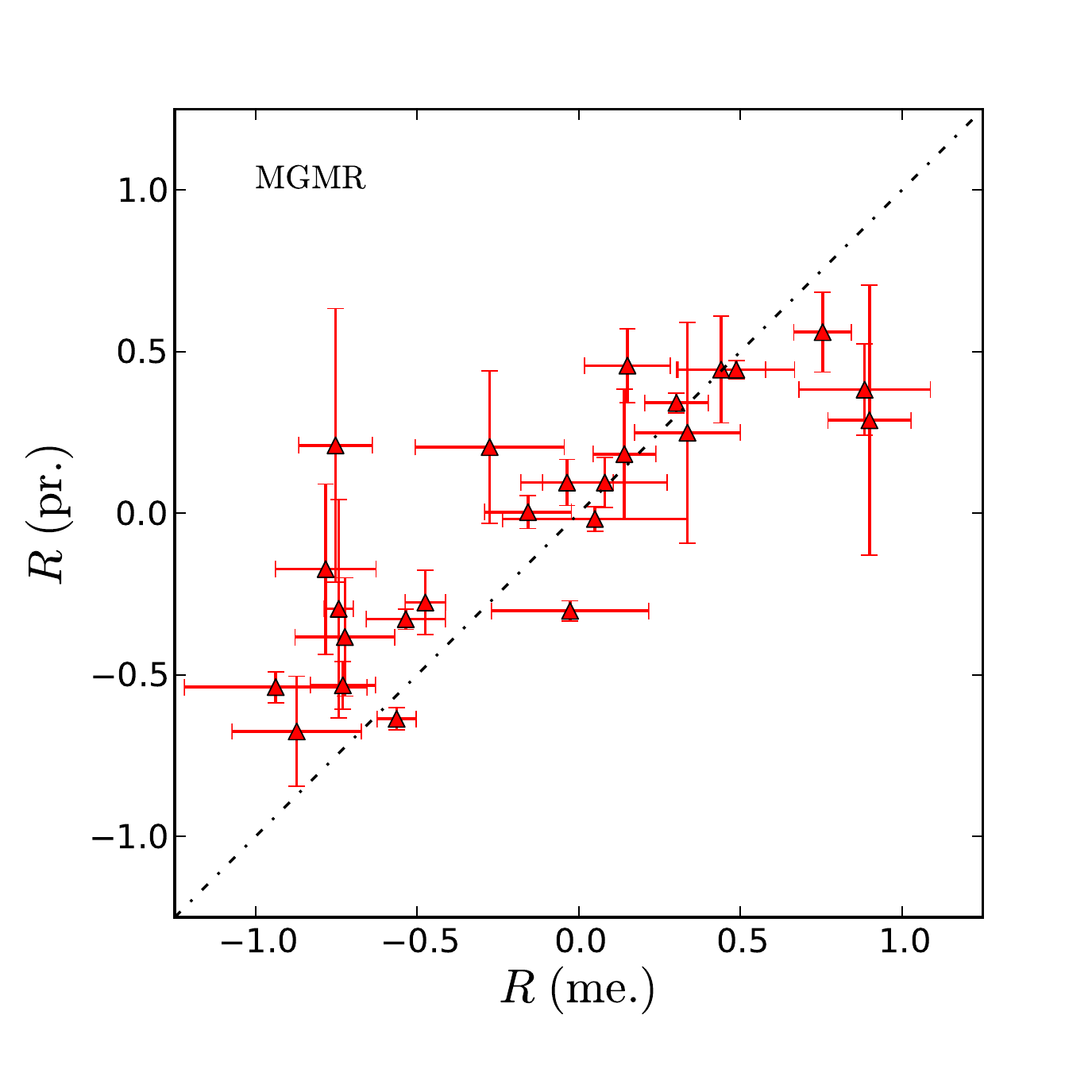}}
    \caption{Measured versus predicted values of the parameter $R$ for the data obtained with the prototype; see also the caption to Fig. \ref{fig:RversusRAERA}.}
	\label{fig:RversusRMAXI}
\end{figure}

In this Appendix we show the results obtained with the prototype. The analysis performed for this data set was essentially the same as the one made for AERA24. In Fig. \ref{fig:Observer-angle-dependenceMAXI} we display the parameter $R$ as function of the observation angle $\psi$. In the left panel of Fig. \ref{fig:polarizationMAXI} we display the predicted polarization angle for pure geomagnetic emission $(a=0)$. For the case where a radial component was added to this geomagnetic emission process, following the procedures outlined in Section \ref{sec:a_analysis}, the data are displayed in the right panel of Fig. \ref{fig:polarizationMAXI}. In this case, the minimum in the reduced $\chi^2$ value is obtained for $a = 0.11 \pm 0.07$. Finally, following the analysis in Section \ref{sec:comparison}, we show in Fig. \ref{fig:RversusRMAXI} the comparison of predicted and measured values for the parameter $R$ and in Table \ref{tab:pearsonMAXI} we list the Pearson correlation coefficients for these data sets.

\begin{table}[h]
\caption{\label{tab:pearsonMAXI} Pearson correlation coefficients $\rho_P$ and their 95\% confidence ranges.}
\begin{ruledtabular}
\begin{tabular}{l  lll lll}
               &\multicolumn{3}{c}{charge excess}      &\multicolumn{3}{c}{no charge excess} \\ \hline
    model       &$\rho_{L}$ &$\rho_P$     &$\rho_{H}$    &$\rho_{L}$  &$\rho_P$     &$\rho_{H}$         \\ \hline
    CoREAS      &0.41   &0.68   &0.86   &      &      &        \\
    EVA1.01     &0.42   &0.68   &0.86   &-0.33  &0.00   &0.34     \\
    MGMR        &0.49   &0.71   &0.87   &-0.29  &0.05   &0.39     \\
    REAS3.1     &0.40   &0.65   &0.82   &       &      &        \\
    SELFAS      &0.45   &0.66   &0.83   &-0.32  &0.03   &0.30     \\
    ZHAireS     &0.43   &0.69   &0.87   &       &      &        \\
\end{tabular}
\end{ruledtabular}
\end{table}

\clearpage

\end{document}